\documentclass{emulateapj}
\usepackage{graphicx}
\usepackage{multirow}
\usepackage{epsfig}
\usepackage{amsmath}
\newcommand{\sfg}{S$^4$G}
\newcommand{\maggie}{AB\,mag\,arcsec$^{-2}$}

\shorttitle{Surface photometry of \sfg\ galaxies}
\shortauthors{Mu\~{n}oz-Mateos et al.}

\begin{document}

\title{The Spitzer Survey of Stellar Structure in Galaxies (\sfg):
  Stellar Masses, Sizes and Radial Profiles for 2352 Nearby Galaxies.}

\author{Juan Carlos Mu\~noz-Mateos\altaffilmark{1,2},
Kartik Sheth\altaffilmark{2},
Michael Regan\altaffilmark{3},
Taehyun Kim\altaffilmark{4,2,1},
Jarkko Laine\altaffilmark{5},
Santiago Erroz-Ferrer\altaffilmark{6,7},
Armando Gil de Paz\altaffilmark{8},
Sebastien Comer\'on\altaffilmark{9},
Joannah Hinz\altaffilmark{10},
Eija Laurikainen\altaffilmark{5,9},
Heikki Salo\altaffilmark{5},
E$.$ Athanassoula\altaffilmark{11},
Albert Bosma\altaffilmark{11},
Alexandre Y$.$ K$.$ Bouquin\altaffilmark{8},
Eva Schinnerer\altaffilmark{12},
Luis Ho\altaffilmark{13,14},
Dennis Zaritsky\altaffilmark{15},
Dimitri Gadotti\altaffilmark{1},
Barry Madore\altaffilmark{14},
Benne Holwerda\altaffilmark{16},
Kar\'in Men\'endez-Delmestre\altaffilmark{17},
Johan H$.$ Knapen,\altaffilmark{6,7}
Sharon Meidt\altaffilmark{12},
Miguel Querejeta\altaffilmark{12},
Trisha Mizusawa\altaffilmark{18},
Mark Seibert\altaffilmark{14},
Seppo Laine\altaffilmark{19},
Helene Courtois\altaffilmark{20}
}

\altaffiltext{1}{European Southern Observatory, Casilla 19001, Santiago 19, Chile; jmunoz@eso.org}
\altaffiltext{2}{National Radio Astronomy Observatory / NAASC, 520 Edgemont Road, Charlottesville, VA 22903, USA}
\altaffiltext{3}{Space Telescope Science Institute, 3700 San Martin Drive, Baltimore, MD 21218, USA}
\altaffiltext{4}{Astronomy Program, Department of Physics and
  Astronomy, Seoul National University, Seoul 151-742, Korea}
\altaffiltext{5}{Astronomy Division, Department of Physics, FI-90014 University of Oulu, PO Box 3000, Oulu, Finland}
\altaffiltext{6}{Instituto de Astrofísica de Canarias, Vía Lácteas/n E-38205 La Laguna, Spain}
\altaffiltext{7}{Departamento de Astrofísica, Universidad de La Laguna, E-38206 La Laguna, Spain}
\altaffiltext{8}{Departamento de Astrofísica, Universidad Complutense de Madrid, E-28040 Madrid, Spain}
\altaffiltext{9}{Finnish Centre of Astronomy with ESO (FINCA), University of Turku, V\"ais\"al\"antie 20, FI-21500 Piikki\"o, Finland}
\altaffiltext{10}{MMTO, University of Arizona, 933 North Cherry Avenue, Tucson, AZ 85721, USA}
\altaffiltext{11}{Aix Marseille Universite, CNRS, LAM (Laboratoire d'Astrophysique de Marseille) UMR 7326, F-13388 Marseille, France}
\altaffiltext{12}{Max-Planck-Institut für Astronomie/Königstuhl 17 D-69117 Heidelberg, Germany}
\altaffiltext{13}{Kavli Institute for Astronomy and Astrophysics, Peking University, Beijing 100871, China }
\altaffiltext{14}{The Observatories of the Carnegie Institution of Washington, 813 Santa Barbara Street, Pasadena, CA 91101, USA }
\altaffiltext{15}{Steward Observatory, University of Arizona, 933 North Cherry Avenue, Tucson, AZ 85721, USA }
\altaffiltext{16}{European Space Agency Research Fellow (ESTEC), Keplerlaan, 1, 2200 AG Noordwijk, The Netherlands}
\altaffiltext{17}{Universidade Federal do Rio de Janeiro, Observat\'orio
  do Valongo, Ladeira Pedro Ant\^onio, 43, CEP20080-090, Rio de
  Janeiro, Brazil}
\altaffiltext{18}{Florida Institute of Technology, Melbourne, FL 32901, USA}
\altaffiltext{19}{Spitzer Science Center, Mail Stop 314-6, 1200 East
  California Boulevard, Pasadena, CA 91125, USA}
\altaffiltext{20}{Institut de Physique Nucl\'{e}aire, Universit\'{e} Lyon 1, CNRS/IN2P3, F-69622 Lyon, France}

\altaffiltext{}{}

\begin{abstract}
  The Spitzer Survey of Stellar Structure in Galaxies (\sfg) is a
  volume, magnitude, and size-limited survey of 2352 nearby galaxies
  with deep imaging at 3.6 and 4.5\,$\micron$. In this paper we describe
  our surface photometry pipeline and showcase the associated data
  products that we have released to the community. We also identify
  the physical mechanisms leading to different levels of central
  stellar mass concentration for galaxies with the same total stellar
  mass. Finally, we derive the local stellar mass-size relation at
  3.6\,$\micron$ for galaxies of different morphologies. Our radial
  profiles reach stellar mass surface densities below $\sim
  1\,M_{\sun}\,\mathrm{pc}^{-2}$. Given the negligible impact of dust
  and the almost constant mass-to-light ratio at these wavelengths,
  these profiles constitute an accurate inventory of the radial
  distribution of stellar mass in nearby galaxies. From these profiles
  we have also derived global properties such as asymptotic magnitudes
  (and the corresponding stellar masses), isophotal sizes and shapes,
  and concentration indices. These and other data products from our
  various pipelines (science-ready mosaics, object masks, 2D image
  decompositions, and stellar mass maps), can be publicly accessed at
  IRSA ({\tt http://irsa.ipac.caltech.edu/data/SPITZER/S4G/}).
\end{abstract}

\section{Introduction}\label{intro}
Understanding how galaxies acquired their baryons over cosmic time is
a key open question in extragalactic astronomy. How and when did
galaxies of different types assemble the bulk of their stellar mass?
Nearby galaxies are of particular relevance in this context: they
constitute the present-day product of billions of years of evolution,
so the past assembly history of these galaxies is encoded in the
present-day spatial distribution of old stars within them. Therefore,
an accurate census of the current stellar structure of nearby galaxies
provides essential constraints on the physics of galaxy formation and
evolution. To address this critical issue, in this paper we present
deep mid-infrared radial profiles for the more than 2300 galaxies in
the Spitzer Survey of Stellar Structure in Galaxies (\sfg, see
\citealt{Sheth:2010} for the full survey description).

Large surveys of nearby galaxies traditionally have been carried out
in the optical regime, for example the Third Reference Catalogue of
Bright Galaxies (RC3, \citealt{de-Vaucouleurs:1991}), the Sloan Digital
Sky Survey (SDSS, \citealt{York:2000}), or the Carnegie-Irvine Galaxy
Survey (\citealt{Ho:2011}). However, translating optical measurements
into stellar masses is not a straightforward task. On the one hand,
the optical mass-to-light ratio ($M_{\star}/L$) is strongly dependent on the
star formation history of the galaxy (e.g$.$ \citealt{Bell:2001}). On
the other hand, internal extinction by dust obscures a significant
fraction of the optical output of galaxies (e.g$.$,
\citealt{Calzetti:2001}). Observations at near- and mid-IR bands can
circumvent these issues. At these long wavelengths, $M_{\star}/L$ is only a
shallow function of the star formation history, and dust extinction
plays a minor role, leading to milder variations in $M_{\star}/L$ compared to
optical wavelengths (\citealt{Meidt:2014}).

Several wide field IR surveys have been carried out over the years,
such as the Two Micron All Sky Survey (2MASS,
\citealt{Skrutskie:2006uq}, see also
\citealt{Jarrett:2000,Jarrett:2003}), the Deep Near Infrared Survey
(DENIS, \citealt{Epchtein:1994}) and more recently the Wide-field
Infrared Survey Explorer (WISE, \citealt{Wright:2010}). These surveys
provide robust number statistics, but at the expense of a shallow
image depth, insufficient to map the stellar content in the faint
outskirts of galaxies. In the case of WISE, the spatial resolution at
3.4 and 4.6\,$\micron$ is relatively coarse ($\sim
6$\arcsec). Conversely, other surveys have obtained much deeper IR
images for small samples of several tens or a few hundred
objects. Projects like these include the Ohio State University Bright
Galaxy Survey (OSUBGS, \citealt{Eskridge:2002}), the Near-Infrared
atlas of S0-Sa galaxies (NIRS0S, \citealt{Laurikainen:2011}), the Spitzer
Infrared Nearby Galaxies Survey (SINGS, \citealt{Kennicutt:2003}) and
the Spitzer Local Volume Legacy Survey (LVL, \citealt{Dale:2009}). But
a complete understanding of galaxy assembly involves many independent
parameters, such as galaxy mass, morphology, environment, bar
presence, gas and dark matter content, etc. With so many independent
dimensions along which a sample should be sliced, any sample with a
few hundreds objects will be broken down into bins too small for a
reliable understanding of the impact of any given parameter.

\sfg\ was specifically designed to answer the need for a deep, large
and uniform IR survey of nearby galaxies. The survey contains over
2352 galaxies within 40\,Mpc, away from the galactic plane
($|b|>30^{\circ}$), with an extinction-corrected B-band Vega magnitude
brighter than 15.5 and a B-band diameter larger than 1\arcmin. We
observed these galaxies at 3.6 and 4.5\,$\micron$ with the Infrared
Array Camera (IRAC, \citealt{Fazio:2004}) onboard Spitzer
(\citealt{Werner:2004}). We followed the successful observing strategy
of the SINGS and LVL programs, reaching azimuthally-averaged stellar
mass surface densities $< 1\,M_{\sun}\,\mathrm{pc}^{-2}$;
this is a physical regime where the baryonic mass budget is dominated
by atomic gas.

The \sfg\ images are processed through a suite of different pipelines
designed to produce a wealth of enhanced data products. The first
three pipelines are described in this paper. Pipeline 1 (P1) creates
science-ready mosaics by combining all the individual exposures of
each galaxy. Pipeline 2 (P2) masks out foreground stars, background
galaxies and artifacts. Pipeline 3 (P3) performs surface photometry on
the images and derives integrated quantities such as asymptotic
magnitudes, isophotal sizes, etc. Pipeline 4 (P4, \citealt{Salo:2015})
uses GALFIT (\citealt{Peng:2002, Peng:2010}) to decompose each galaxy
into bulges, disks, bars, etc. Finally, Pipeline 5 (P5,
\citealt{Querejeta:2014}) combines the 3.6 and 4.5\,$\micron$ images
to produce stellar mass maps via an Independent Component Analysis,
following the methodology developed by \cite{Meidt:2012}.

The goal of this paper is twofold. First, we describe the technical
details and inner workings of P1, P2 and P3, as well as the resulting
data products. Then, we discuss two particular scientific applications
of our data: the local stellar mass-size relation, and the physics
behind the vast diversity of morphologies in galaxies with the same
stellar mass.

For each galaxy we have obtained radial profiles both with fixed and
free ellipticity and position angle. From these profiles we have also
derived global measurements such as asymptotic magnitudes and stellar
masses, isophotal sizes and ellipticities, and concentration
indices. This dataset constitutes a unique tool to address many
important issues on stellar structure, including but not limited to:

\begin{itemize}
\item the radial structure of dark matter in galaxies. Our profiles
  constrain the radial distribution of stellar mass, which is a
  necessary ingredient when modelling rotation curves to infer the
  radial distribution of dark matter (\citealt{Bosma:1978, Rubin:1978,
    Sofue:2001, de-Blok:2008}).
\item the scaling laws of disks (\citealt{Courteau:2007}), in
  particular the local stellar mass-size relation
  (\citealt{Kauffmann:2003, Shen:2003}) and, when kinematic
  data are available, the Tully-Fisher relation (\citealt{Tully:1977, Aaronson:1979,
    Verheijen:2001, Sorce:2012, Zaritsky:2014}).
\item the inside-out assembly of disks, by comparing our profiles of
  old stars with UV and optical profiles probing younger stellar populations
  (\citealt{de-Jong:1996, Bell:2000, MacArthur:2004, Taylor:2005,
    Munoz-Mateos:2007, Munoz-Mateos:2011, Wang:2011}).
\item the local bar fraction and the sizes, strengths and shapes of bars
  (\citealt{Eskridge:2000, Knapen:2000, Whyte:2002, Erwin:2005a,
    Menendez-Delmestre:2007, Marinova:2007, Sheth:2008, Kim:2014}).
\item radial migration of old stars and the assembly of galactic
  outskirts (\citealt{Roskar:2008, Sanchez-Blazquez:2009,
    Minchev:2011, Martin-Navarro:2012}), in particular the study of
  breaks in radial projected surface density profiles and their links
  to bar resonances (\citealt{Munoz-Mateos:2013}).
\item one-dimensional structural decompositions
  (\citealt{Baggett:1998,MacArthur:2003}), from which one can derive
  physical parameters such as disk scale-lengths, bulge effective
  radii, bulge-to-disk ratios, etc.
\item the intrinsic face-on circularity of disks and their vertical
  thickness (\citealt{Sandage:1970, Ryden:2004, Ryden:2006,
    Comeron:2011, Zaritsky:2013}).
\item quantified galaxy morphology via non-parametric estimators such
  as concentration indices (\citealt{Bershady:2000, Conselice:2003,
    Abraham:2003, Munoz-Mateos:2009a, Holwerda:2013, Holwerda:2014}).
\item the internal structure of elliptical galaxies
  (\citealt{Kormendy:2009}), in particular their intrinsic axial
  ratios (\citealt{Ryden:1992}), their S\'{e}rsic indices
  (\citealt{Caon:1993}) and their boxy/disky structure
  (\citealt{Bender:1988, Peletier:1990}).

\end{itemize}

This paper is structured as follows. In Section~\ref{sec_sample} we
summarize our sample selection criteria and
observations. Section~\ref{sec_P1} contains a brief outline of the
image reduction performed by P1. A more detailed description can be
found in Appendix~\ref{App_P1}. The object masking procedure is
described in Section~\ref{sec_P2}. Then, in Section~\ref{sec_P3} we
explain our ellipse fitting technique and the resulting data
products. Readers interested in the scientific applications of these
products can proceed to Section~\ref{sec_results}, where we discuss
the stellar-mass size relation, as well as the variety of morphologies
and radial concentration that we find at a fixed stellar mass. Then,
Section~\ref{sec_irsa} explains how to access and download our
data. Finally, in Section~\ref{sec_summary} we summarize our main
conclusions.

\section{Sample and observations}\label{sec_sample}
To assemble the \sfg\ sample we made use of the HyperLEDA database
(\citealt{Paturel:2003}). We selected all galaxies with radio-derived
radial velocity $v<3000$\,km/s, which translates into a distance cut
of $d \lesssim 40$\,Mpc for $H_0 = 71$\,km\,s$^{-1}$\,Mpc$^{-1}$. We
only considered galaxies with total corrected blue magnitude
$m_{B_{\mathrm{corr}}}<15.5$, blue light isophotal diameter
$D_{25}>1.0$\arcmin, and Galactic latitude $|b|>30^{\circ}$. The final
sample after applying all these selection criteria comprises 2352
galaxies. Using radio-based velocities biases the sample against
gas-poor early-type galaxies. However, our recently approved Cycle 10
program (prog. ID 10043, PI: K. Sheth) will complete Spitzer's legacy
with archival and new observations of all $\sim 700$ early-type
galaxies within the \sfg\ volume, following the same observing
strategy and selection cuts.

Out of the 2352 \sfg\ galaxies, $\sim 25$\% were already present in
the Spitzer archive as a result of previous programs carried out
during the cryogenic phase. We observed the remaining $\sim 75$\%
during the post-cryogenic mission. In this regard, from now on we will
refer to the galaxies in our sample as either ``archival'' or
``warm'', respectively.

We observed the warm galaxies with a total on-source exposure time of
240\,s per pixel, and mapped them out to at least $1.5 \times
D_{25}$. Depending on the apparent size of each galaxy, we used either
a single dithered map or a mosaic of several pointings. Each galaxy
was observed in two visits separated by at least 30 days, in order to
image the galaxy with two different orientations thanks to the
rotation of the telescope. This allowed for a better correction of
cosmetic effects, cosmic ray and asteroid removal, and subpixel
sampling.

Most of the archival galaxies had been also mapped with at least
240\,s per pixel and out to $1.5 \times D_{25}$.  However, six
  archival galaxies had exposure times between 90 and 200\,s per
  pixel, and 125 were mapped out to less than $1.5 \times
  D_{25}$. They represent a small fraction of the total sample, and
  only specific science goals are affected by this. We therefore
  decided not to reobserve these galaxies, as the incremental gain of
  repeating these observations would not have made up for the required
  additional observing time (see \citealt{Sheth:2010} for more details
  on the sample selection).

\section{Pipeline 1: Science-ready images}\label{sec_P1}
We refer the reader to Appendix~\ref{App_P1} for a more detailed
description of P1. Briefly speaking, P1 creates science-ready mosaics
by combining the different exposures of each galaxy. The pipeline
first matches the background level of the individual exposures, using
the overlapping regions among them. It then combines all frames
following standard dither/drizzle procedures
(\citealt{Fruchter:2002}). The final science-ready mosaics are
delivered in units of MJy\,sr$^{-1}$, with a pixel scale of
0.75\arcsec. The FWHM of the PSF at 3.6 and 4.5\,$\micron$ is
1.7\arcsec\ and 1.6\arcsec, respectively. For the farthest galaxies in
our sample at $\sim 40$\,Mpc, this translates into a physical size of
$\sim 300$\,pc.

Besides the scientific images themselves, P1 also produces weight-maps
showing how many individual frames cover each pixel of the final
mosaics.

\section{Pipeline 2: Object masks}\label{sec_P2}
For each galaxy in the sample, P2 creates masks of contaminant sources
like background galaxies, foreground stars and artifacts. Such masks
are necessary for subsequent analysis such as background measurements
and surface photometry (P3), 2D image decompositions (P4) and mass
map generation (P5).

The mask creation procedure consists of two steps: (i) automatic
generation of initial masks, and (ii) visual check and editing to
create final clean masks. A first set of masks is automatically
generated by running SExtractor (\citealt{Bertin:1996}) on the 3.6 and
4.5$\micron$ images; the resulting segmentation maps constitute our
initial raw masks. Masking sources is relatively easy in those areas
of an image with little or no emission from the main target
galaxy. However, it becomes more complicated on top of the galaxy,
where we need to avoid masking regions belonging to the galaxy
itself. Therefore, for each galaxy and band we create three automatic
masks with high, medium, and low detection thresholds in SExtractor
that control how aggressively different sources are masked on the main
body of the galaxy.

\begin{figure}
\begin{center}
\resizebox{1\hsize}{!}{\includegraphics{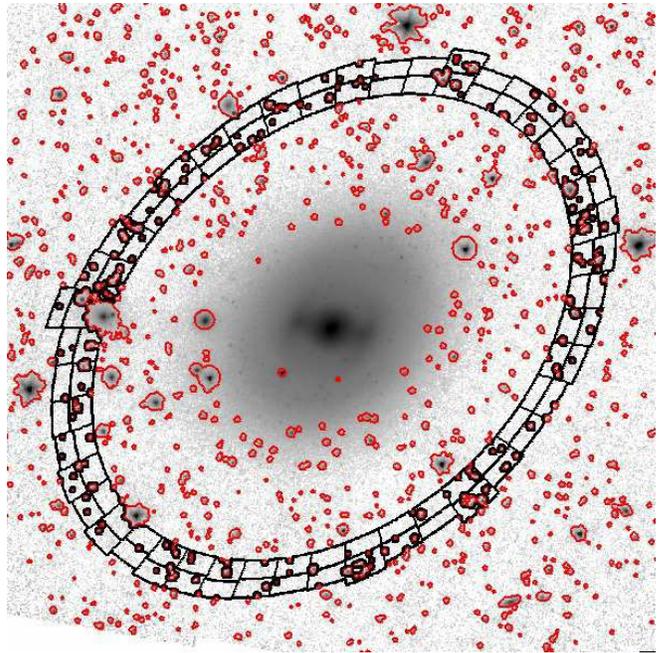}}
\caption{Sample object mask (red) and sky boxes (black) overlaid on
  the 3.6\,$\micron$ image of NGC0936. Note that the sky boxes are
  automatically grown in the radial direction to ensure that they all
  contain the same amount of unmasked pixels.\label{sample_mask}}
\end{center}
\end{figure}

During the visual quality check step, for each galaxy we first choose
the best mask among the three ones with different thresholds. This
best mask is then visually inspected and edited by hand, masking
additional sources missed by SExtractor and unmasking any regions of
the galaxy that may have been masked by mistake. In particular, we
often have to manually add to the masks the extended haloes and
diffraction spikes of bright stars. Conversely, certain sources on the
target galaxies like bright clumps or the ansae of bars are sometimes
picked up by SExtractor, and need to be excluded from the masks by
hand. This editing process is performed with a custom code described
in \cite{Salo:2015}. This IDL routine displays side by side the
  original and masked images, and provides several geometric shapes
  that the user can employ to interactively mask or unmask regions.
The editing is first manually done on the 3.6\,$\micron$ images, and
then the editing is automatically transferred to the 4.5$\micron$
ones, checking that artifacts remain properly masked. Thus, we
end up with two final, edited masks for each galaxy, one for each
band. A sample mask is shown in Fig.~\ref{sample_mask}.

The masking process is to some extent subjective, especially for the
faintest sources, due to the lack of color information. Theses masks
are mainly intended to remove extraneous sources that would
significantly contaminate and distort our radial profiles and
integrated magnitudes. Therefore, users interested in specific
structures such as very faint HII regions or globular clusters around
our galaxies should not blindly use our masks without double-checking
the nature of these sources with ancillary multi-wavelength data.

\section{Pipeline 3: Surface photometry}\label{sec_P3}
In this section we describe in detail the inner workings of Pipeline
3. Briefly speaking, P3 first determines the background level of the
images. It then locates the centroid of the galaxy and performs
different sets of ellipse fits. Finally, it measures several
properties such as asymptotic magnitudes, isophotal sizes and
concentration indices.

\subsection{Sky measurement}\label{sec_sky}
A careful determination of the background level and noise is essential
to perform reliable surface photometry. This is often done by
placing several ``sky boxes'' around each galaxy and measuring the
corresponding background statistics in them. With large samples such
as \sfg\ it is desirable to implement this process in a way that is
automatic yet flexible.

Pipeline 3 automatically places several sky boxes around each galaxy
(see Fig.~\ref{sample_mask}). This is done by defining two concentric
and adjacent elliptical annuli that surround the entire galaxy. Each
ring is then azimuthally subdivided into 45 sectors or boxes. Since
these boxes will contain in general a certain amount of masked pixels,
each box is grown radially outwards until each one contains 1000
unmasked pixels. We then measure the median sky level and local rms
inside each box, as well as the large-scale rms between the sky levels
of all boxes. While the number of 1000 pixels per box is to some
extent arbitrary, it is chosen to provide a reliable measurement of
the local rms, while at the same time yielding boxes that are small
enough to be easily accommodated within our images. Note also that
before using the P2 masks, we first grow the masked areas by 2 pixels
to make sure that the faint wings of the PSF do not contaminate our
sky measurements.

The sky boxes are initially placed by default at $2 \times R_{25}$
from the galaxy center, but this value is modified as needed for each
galaxy in order to ensure a proper background subtraction. To do
this, we compare the values between the inner and outer rings to make
sure that there are no significant differences, which could be a
telltale sign of contamination from the galaxy itself. We double check
this by verifying that the growth curve is flat (see
Section~\ref{sec_magnitudes}). We also check that the boxes are
  not too close to the frame edges (which are noisier as a result of
  the dithering pattern), or that they do not fall in the adjacent
  frames (which may have a somewhat different background value). In
cases with a complicated background structure, the pipeline allows one
to manually distribute the sky boxes as deemed appropriate.

\begin{figure}
\begin{center}
\resizebox{1\hsize}{!}{\includegraphics{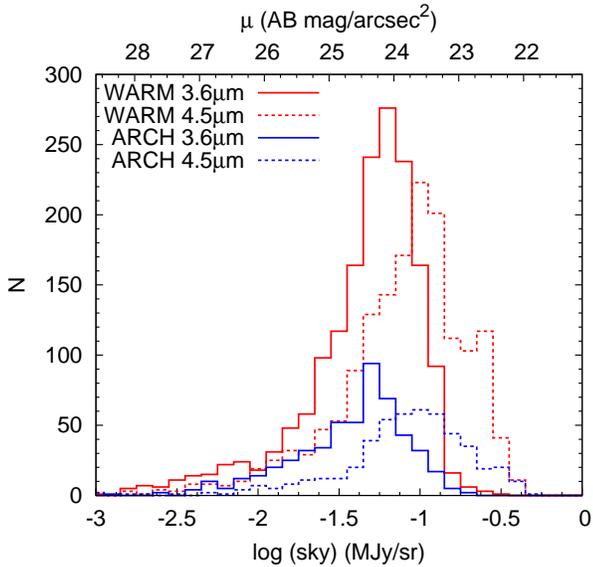}}
\caption{Distribution of the sky level within the full \sfg\
  sample. Red and blue histograms show galaxies from the warm and
  archival (i.e$.$ cryo) groups, respectively. Solid lines correspond
  to 3.6\,$\micron$, and dashed ones to 4.5\,$\micron$.\label{sky_hist}}
\end{center}
\end{figure}

The distribution of measured background levels in the whole \sfg\
sample is shown in Fig.~\ref{sky_hist}. The histograms peak at $\mu
\sim 24$ and $23.5$\,\maggie\ for the 3.6 and 4.5\,$\micron$ bands,
respectively.

It is illustrative to verify whether our measured values of the
background level agree with theoretical expectations. The background
in the \sfg\ images is almost entirely dominated by zodiacal light (or
``zodi'') coming from dust grains in the ecliptic plane. Both the
thermal emission of the grains and the scattered sunlight contribute
to the zodi. At 3.6 and 4.5\,$\micron$, both contributions are roughly
equal. Thermal emission from interstellar cirrus in the Milky Way
amounts to merely 1\%-3\% of the background in our images, as it peaks
at much longer wavelengths (and the \sfg\ sample specifically avoids
the Galactic plane anyway). Contamination by the unresolved Cosmic
Infrared Background is negligible at these bands compared to the zodi
(\citealt{Hauser:1998}).

\begin{figure}
\begin{center}
\resizebox{1\hsize}{!}{\includegraphics{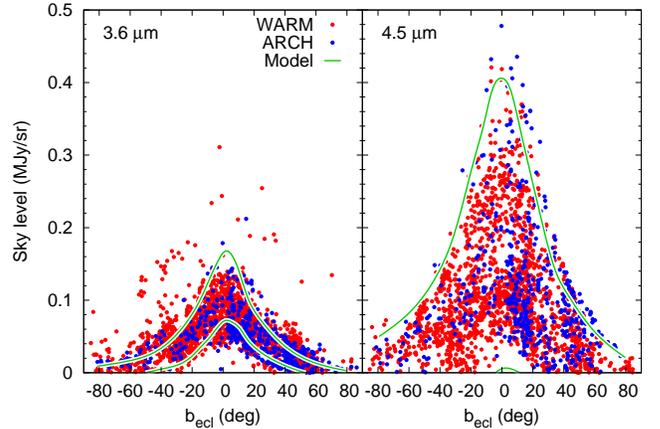}}
\caption{Background level measured in our fully-reduced images as a
  function of the ecliptic latitude of each galaxy. Red and blue dots
  correspond to warm and archival galaxies. The green curves delimit
  the distribution of the zodi brightness predicted by the Spitzer
  background estimator.\label{sky_ecliptic}}
\end{center}
\end{figure}

Figure~\ref{sky_ecliptic} shows how the observed sky level varies with
the ecliptic latitude of each galaxy. As expected, since the primary
source of background emission at these wavelengths is zodi, the
distribution clearly peaks in the ecliptic plane. For each galaxy we
retrieved from the FITS header the background level predicted for its
Galactic coordinates and epoch of observation.\footnote{For details on
  the Spitzer background estimator, see\\{\tt
    http://ssc.spitzer.caltech.edu/warmmission/propkit/som/bg/}}$^{,}$\footnote{The
  predicted zodi at the date and coordinates of observation is stored
  in the FITS header keyword {\tt ZODY\_EST}. Part of this zodi is removed
  when the Spitzer automatic pipeline subtracts a skydark frame; the
  estimated amount of removed zodi is given by {\tt SKYDRKZB}. The
  final level of zodi that remains in our images is therefore {\tt
    ZODY\_EST} $-$ {\tt SKYDRKZB}.} The green
curves in Fig.~\ref{sky_ecliptic} are the upper and lower envelopes of
these predicted values. We can see that both the dependency with
ecliptic latitude and the range in background level at fixed latitude
nicely agree with our measured values.

\begin{figure}
\begin{center}
\resizebox{1\hsize}{!}{\includegraphics{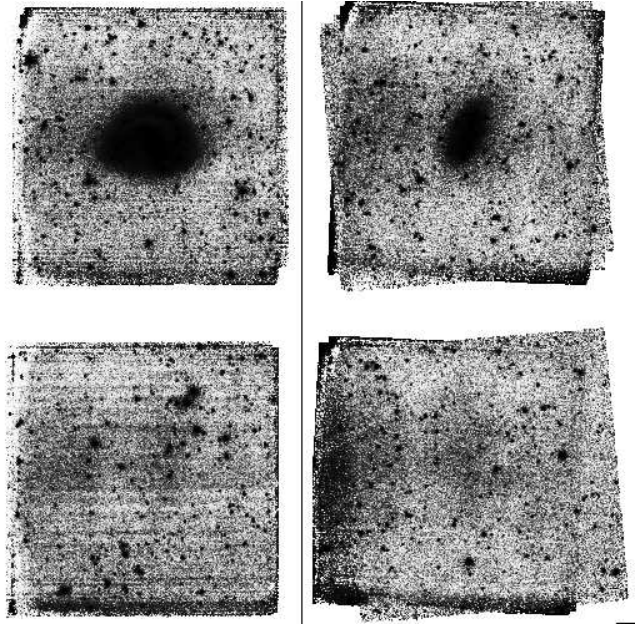}}
\caption{Two sample images exhibiting a diffuse background artifact:
  NGC5597 (left) and UGC08507 (right). The smooth pattern is seen
  both in the main frame (top) and in the flanking one (bottom). A
  histogram equalization has been applied to exaggerate the
  brightness of this artifact. This structure only appears in $\sim
  30$ of our galaxies.\label{backgr_artifacts}}
\end{center}
\end{figure}

Interestingly, about 30 WARM galaxies exhibit background levels at
3.6\,$\micron$ well above the predicted values. A visual inspection of
these images revealed a recurrent diffuse artifact, both in the main
frame where the galaxy is and in the flanking one
(Fig.~\ref{backgr_artifacts}). These few galaxies were observed in a
narrow time window during August-September 2009, very shortly after
the Spitzer's warm phase began. This smooth background artifact may be
therefore due to the detector temperature and bias not being settled
yet at that time.

\begin{figure}
\begin{center}
\resizebox{0.49\hsize}{!}{\includegraphics{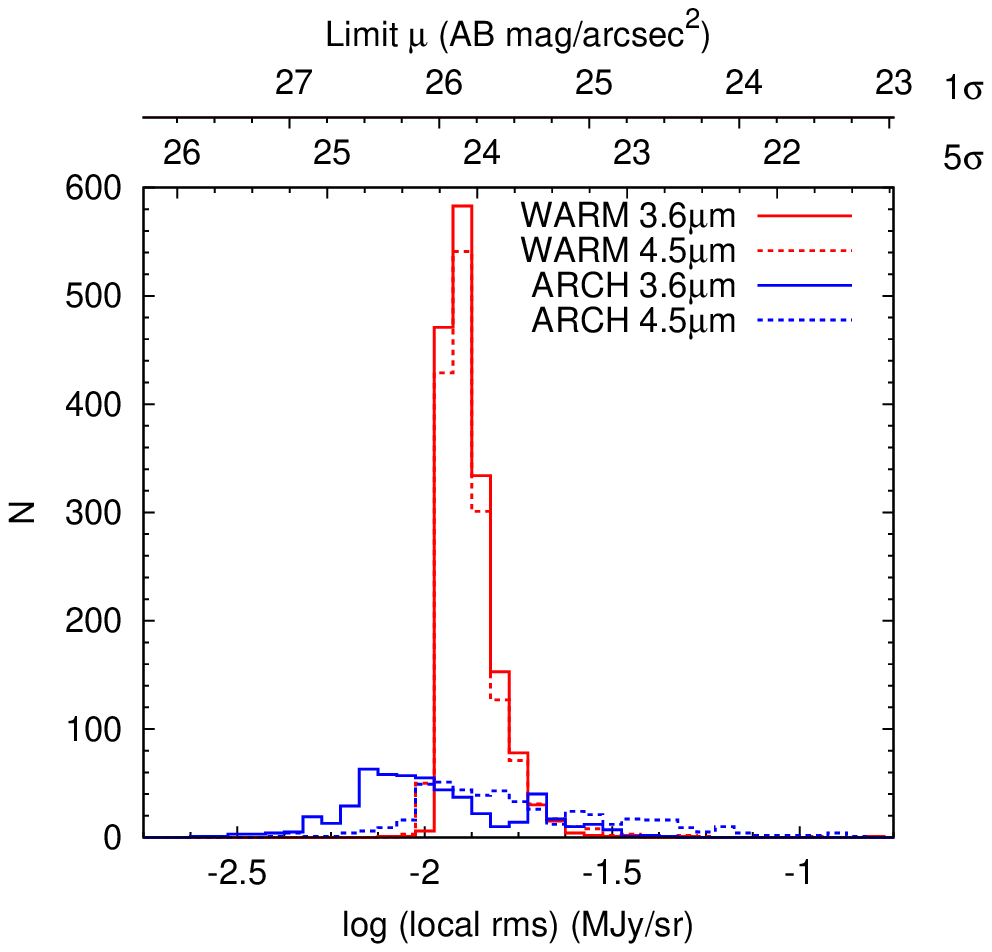}}
\resizebox{0.49\hsize}{!}{\includegraphics{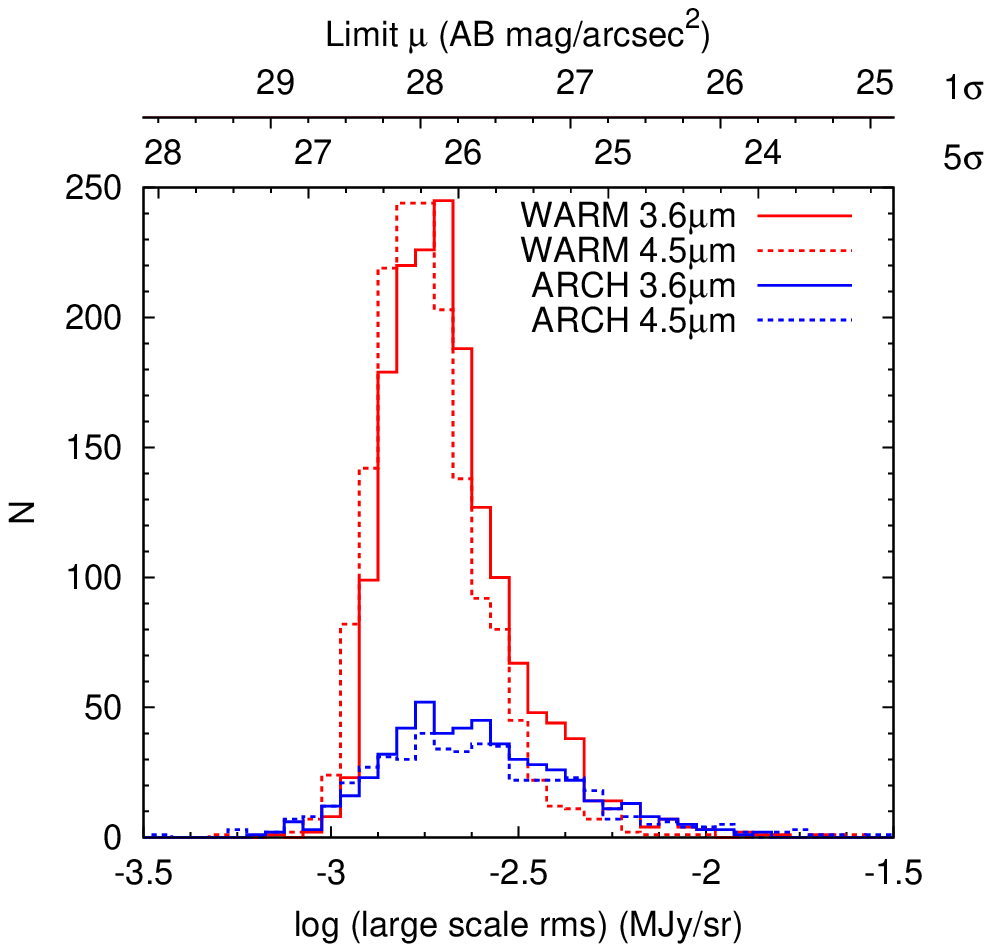}}\\
\caption{Distribution of local, pixel-to-pixel noise (left) and large
  scale background noise (right) in the \sfg\ images. WARM and
  ARCHIVAL galaxies are shown in red and blue, respectively. Solid
  and dashed lines correspond to 3.6 and 4.5\,$\micron$. The
  corresponding surface brightness limit at $5\sigma$ and $1\sigma$
  is shown at the top.\label{sky_noise}}
\end{center}
\end{figure}

Besides the background level itself, it is also important to
characterize the background noise at different spatial scales, as this
determines our ability to detect and measure faint structures at the
outskirts of our galaxies. Figure~\ref{sky_noise} shows the
distribution of the local, pixel-to-pixel noise in our images. The
histogram for the warm galaxies is much narrower than that for the
archival ones, which is expected given that the warm galaxies were
imaged with the same observing strategy, whereas the archival come
from a variety of different programs. In general we reach a local
surface brightness sensitivity per pixel of $\sim 24$ and $\sim
26$\,\maggie\ at $5\sigma$ and $1\sigma$, respectively.

Note, however, that these are values {\it for individual pixels}. When
measuring the flux of a given extended source, this noise component
scales down with the square root of the number of pixels in the region
where the measurement is being performed. In particular, when
measuring radial profiles at the outskirts of galaxies, this local
noise component becomes negligible compared to the large scale
background noise, which is given by the rms between the median sky
values measured in the different boxes. As shown in
Fig.~\ref{sky_noise}, this large scale component peaks at
26-26.5\,\maggie\ at a $5\sigma$ level, and $\sim 28$\,\maggie\ at
$1\sigma$. As a reference, $\mu_{3.6} = 27$\,\maggie\ corresponds to a
stellar mass surface density of $1\,M_{\sun}\,\mathrm{pc}^{-2}$,
adopting $M_{\star}/L_{3.6}=0.53$ as measured by \cite{Eskew:2012};
see also \cite{Meidt:2014}.

\subsection{Radial profiles}
\begin{figure}
\begin{center}
\begin{tabular}{cc}
\resizebox{0.3\hsize}{!}{\includegraphics{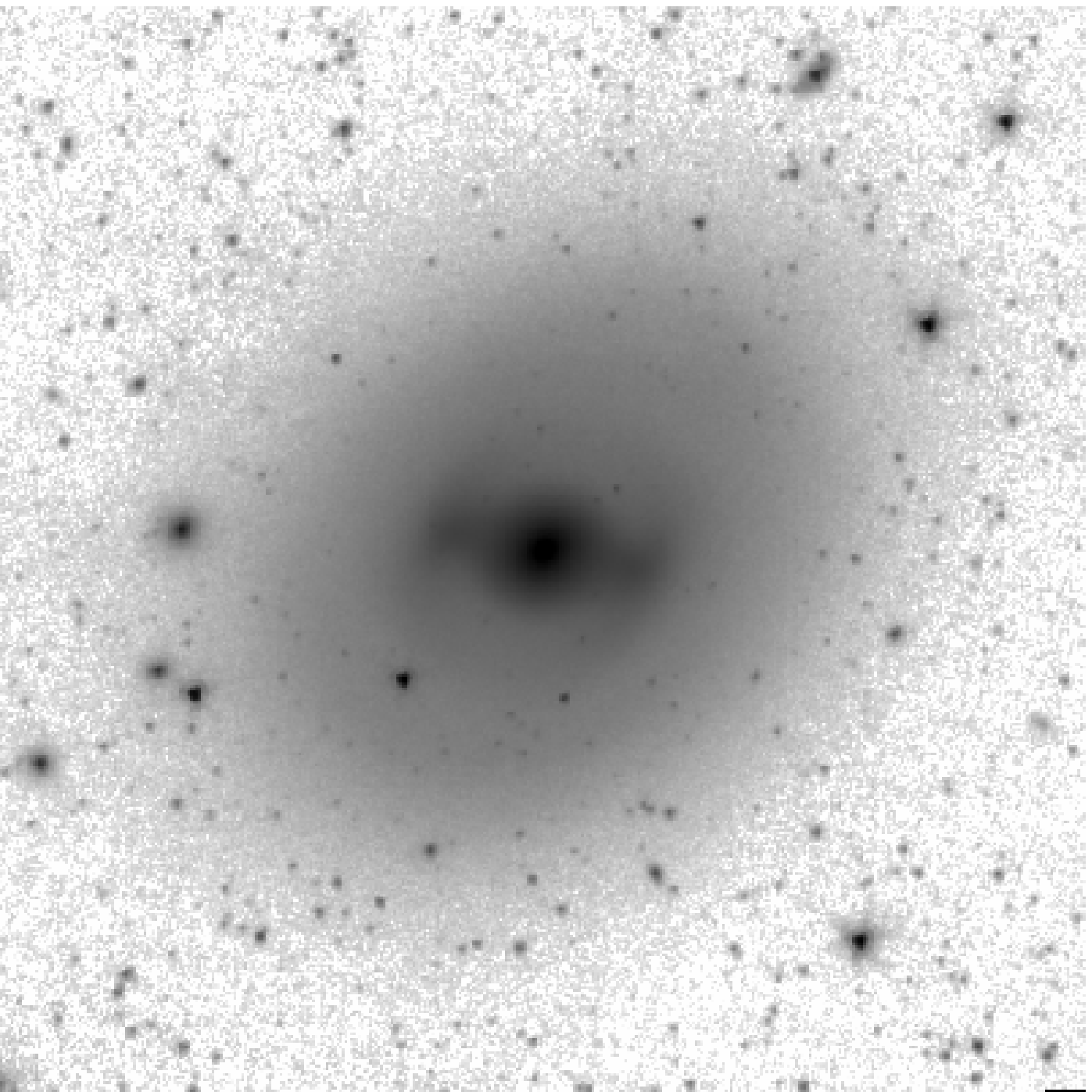}} & \multirow{2}{*}[22ex]{\resizebox{0.53\hsize}{!}{\includegraphics{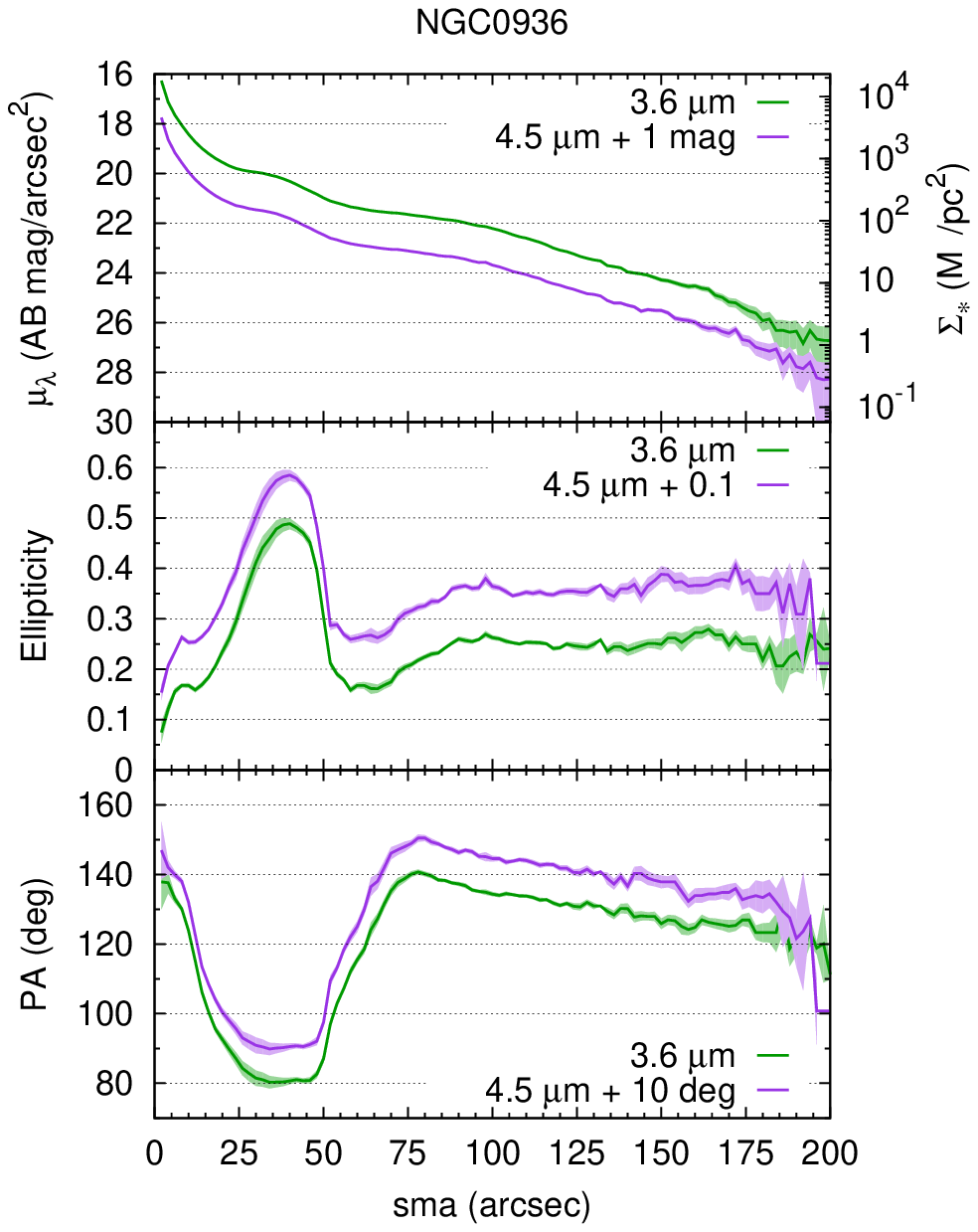}}}\\
\resizebox{0.3\hsize}{!}{\includegraphics{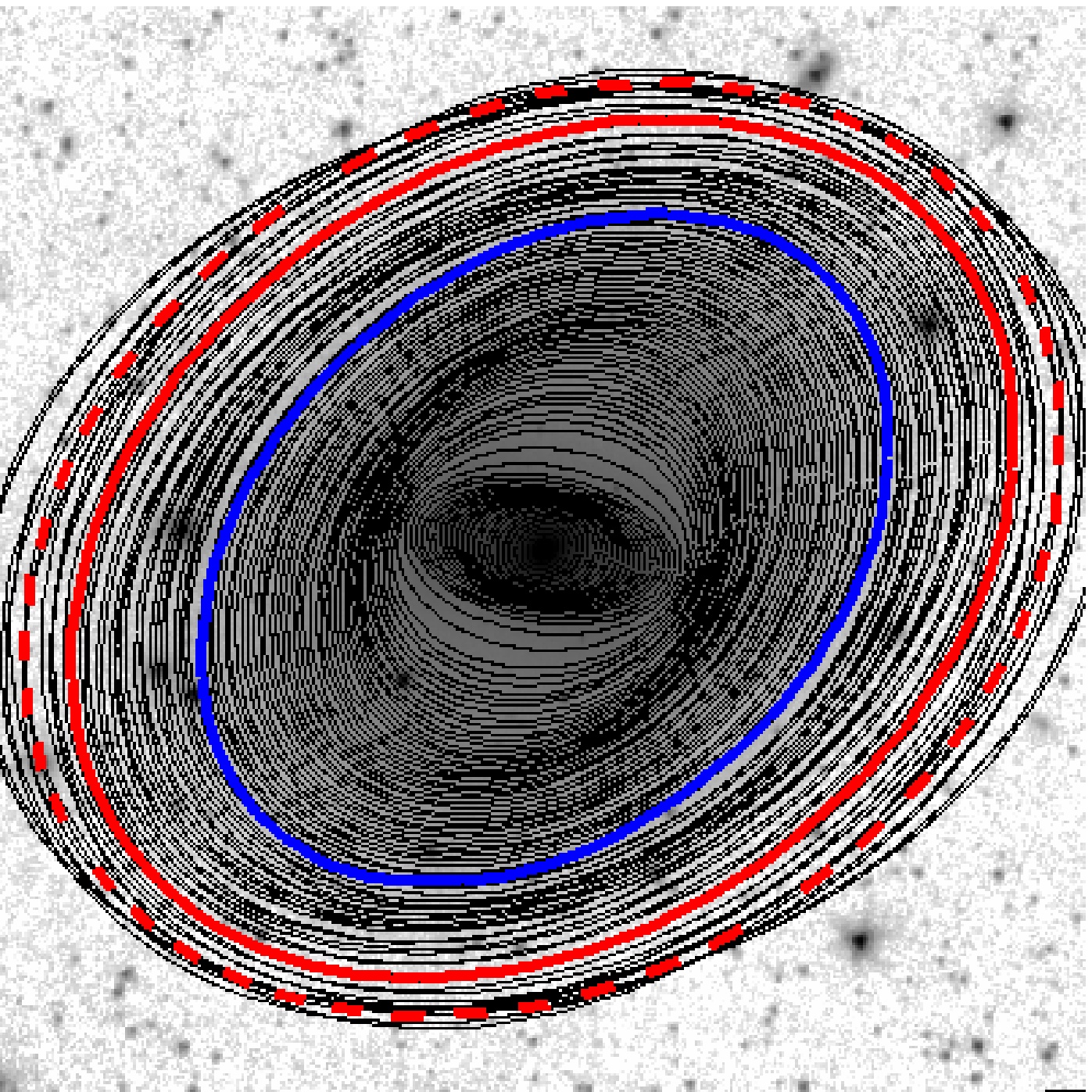}} & \\
\end{tabular}
\caption{Sample ellipse fit results for NGC0936 with free $\epsilon$
  and PA, and a radial increment $\Delta r=2$\arcsec. The images to
  the left correspond to the 3.6\,$\micron$ band. In the bottom one we
  have overlaid the ellipses fitted by our pipeline. The red solid
  ellipse corresponds to $\mu_{3.6} = 25.5$\,\maggie, and the dashed
  one to $26.5$\,\maggie. The blue ellipse is the optical $D_{25}$ one
  from HyperLeda. In the radial profiles, offsets have been applied to
  the 4.5\,$\micron$ data as indicated in the legend, to avoid
  overlapping. The right vertical scale in the top panel marks the
  stellar mass surface density for the 3.6\,$\micron$
  profile.\label{sample_prof}}
\end{center}
\end{figure}

Once the background level and noise have been measured, our pipeline
proceeds to perform surface photometry on the images. We first use the
IRAF\footnote{IRAF is distributed by the National Optical Astronomy
  Observatory, which is operated by the Association of Universities
  for Research in Astronomy (AURA) under cooperative agreement with
  the National Science Foundation.} task {\sc imcentroid} to find
accurate coordinates for the center of the galaxy. We then run the
task {\sc ellipse} (\citealt{Jedrzejewski:1987, Busko:1996}) to obtain
radial profiles of surface brightness ($\mu$), ellipticity
($\epsilon$) and position angle (PA). We perform three separate runs
of {\sc ellipse} with different settings:

\begin{enumerate}
\item {\bf Fixed center; free $\epsilon$ and PA; radial resolution $\Delta
  r=6$\arcsec.} We use these fits with a coarse radial increment to
  derive $\epsilon$ and PA in the outer parts of the galaxy, thus
  defining a global shape and orientation for each object. For each
  band, we provide values of $\epsilon$ and PA at two levels of
  surface brightness, $\mu = 25.5$ and $26.5$\,\maggie. After testing
  how sensitive these values are to variations in the sky subtraction,
  input fitting parameters, amount of masked objects, etc., we
  recommend using the 25.5 values as they are more stable.

\item {\bf Fixed center; $\epsilon$ and PA fixed to the values at
  25.5\,\maggie; $\Delta r=2$\arcsec.} These fits have a finer radial
  resolution that matches the IRAC PSF at these wavelengths. By
  keeping $\epsilon$ and PA fixed and equal to the global outer
  values, these fixed-fits are ideal to measure disk scale-lengths,
  disk break radii, to perform 1D bulge-disk decompositions, etc. We
  also employ these profiles to measure the integrated magnitude of
  each galaxy from the growth curve (Sect.~\ref{sec_magnitudes}).

\item {\bf Fixed center; free $\epsilon$ and PA; $\Delta r=2$\arcsec.}
  These free-fits with a fine resolution are well suited to study in
  detail the structural properties of features such as bars, which
  leave very characteristic signatures in the radial profiles of
  $\epsilon$ and PA. An example of these fits is shown in
  Fig.~\ref{sample_prof}.

\end{enumerate}

For simplicity and to improve the robustness of the fits, once we
measure the central coordinates of galaxy, these are kept fixed
during the ellipse fitting. However, the pipeline allows one to leave
the center as a free parameter too. We have in fact performed such
fits to study individual galaxies with offset bars (e.g$.$ NGC3906, de
Swardt et al$.$ 2014, in prep.).

We normally begin the ellipse fitting at an intermediate radius,
typically between $0.5 - 1\times R_{25}$, using the optical $\epsilon$
and PA from HyperLEDA as input guesses for the fitting
routine. Starting at that initial radius, the ellipse fitting proceeds
first outwards and then inwards.

It should be noted that the {\sc ellipse} algorithm was specifically
designed for galaxies with a smooth radial brightness distribution. It
therefore works correctly for elliptical galaxies, but it does not
cope well with features such as spiral arms or rings that can
significantly modify the local luminosity gradient. As a result, {\sc
  ellipse} can sometimes stop fitting at a given radius in spiral
galaxies. Tuning the input parameters can sometimes fix this problem
and provide a good fit, but in general this is a blind trial-and-error
solution with an unpredictable outcome. Some authors have in fact
circumvented this issue by automatically running {\sc ellipse} tens or
even hundreds of times on each image, varying the initial fitting
parameters each time until a good fit is obtained
(\citealt{Jogee:2004}). While this approach can be practical in small
images of distant galaxies, it is prohibitively time-consuming for the
large images in \sfg, where a single fit can take up to several
minutes (and this is further complicated by the large number of
galaxies in our sample). We thus decided to implement an optimized
procedure that is better tailored to our needs. After each run of {\sc
  ellipse}, the pipeline looks for any radial interval where the fit
did not converge, and it then refits that particular radial interval
alone. The process is then iterated several times to ensure that no
radial gap is left unfit. In general most of our galaxies can be
properly fit with a single {\sc ellipse} run, but this automatic
iterative process was of great help to handle troublesome cases that
would have otherwise required considerable manual work.

In any case, for each galaxy we always visually inspect the output of
the pipeline: we overlay the fitted ellipses on the galaxy image, and
we plot the radial profiles of $\mu$, $\epsilon$ and PA, making sure
that the results accurately probe the different structures within the
galaxy.

Dwarf galaxies can be particularly challenging for our pipeline. They
are often partially resolved into stars in our images, and their
clumpy and patchy appearance can sometimes fool the ellipse fitting
algorithm. For these galaxies we recommend not to overinterpret any
signature in the $\epsilon$ and PA profiles without inspecting the
images themselves. Since these galaxies normally lack well-defined
large-scale structures such as bars, the profiles with fixed
$\epsilon$ and PA are more suitable in these cases.

We have released all these ellipse fits in the form of ASCII tables
containing the full output of the {\sc ellipse} task. These tables not
only include radial profiles of $\mu$, $\epsilon$ and PA, but also
other quantities such as the harmonic deviations from a perfect
ellipse, which are commonly used to quantify the boxiness/diskiness of
the isophotes at different radii (e.g$.$ \citealt{Carter:1978,
  Kormendy:1996}).

Apart from the usual output from {\sc ellipse}, we also include in our
ASCII profiles additional columns that are specific to the IRAC data
used here. In particular, the IRAC photometry needs to be corrected
for the extended wings of the PSF and the diffuse light that is
scattered throughout the detector. Here we rely on the extended source
aperture correction provided in the IRAC Instrument
Handbook\footnote{\tt
  http://irsa.ipac.caltech.edu/data/SPITZER/docs/irac/iracinstrumenthandbook/30/}. Given
an elliptical aperture with major and minor radii $a$ and $b$, if
$F_{\mathrm{obs}}$ is the total observed flux inside such an aperture,
the corrected flux is given by:
\begin{equation}
F_{\mathrm{corr}} (r_{\mathrm{eq}}) = F_{\mathrm{obs}} (r_{\mathrm{eq}}) \times (A e^{-r_{\mathrm{eq}}^B}+C)\label{eq_F_aper_corr}
\end{equation}
In this expression $r_{\mathrm{eq}}=\sqrt{ab}$ is the equivalent
radius of the elliptical aperture, in arcseconds. The coefficients
$A$, $B$ and $C$ are equal to 0.82, 0.370 and 0.910, respectively, at
3.6\,$\micron$, and 1.16, 0.433 and 0.94 at 4.5\,$\micron$.

Similarly, if $I_{\mathrm{obs}}$ is the surface brightness along an
isophote (rather than the total flux inside that radius as before), the
aperture-corrected surface brightness can be obtained by performing a
series expansion on the previous equation:
\begin{align}
I_{\mathrm{corr}} (r_{\mathrm{eq}}) = &I_{\mathrm{obs}}(r_{\mathrm{eq}}) \times (A e^{-r_{\mathrm{eq}}^B}+C) - \nonumber\\
&- ABr_{\mathrm{eq}}^{B-2}e^{-r_{\mathrm{eq}}^B} F_{\mathrm{obs}} (r_{\mathrm{eq}})/(2 \pi) \label{eq_I_aper_corr}
\end{align}

Our ASCII tables include radial profiles with and without these
aperture corrections. These corrections are estimated to be uncertain
at a 5\%-10\% level. We have always applied the extended source aperture
correction at all radii in our profiles. However, the reader should
keep in mind that the point source corrections might be more suitable
at very small radii ($r \lesssim 8$-$9$\arcsec), especially if the
nucleus is bright and compact.

Finally, all the released profiles have been corrected for foreground
extinction, using the color excess map of \cite{Schlegel:1998} and the
Milky Way extinction curve of \cite{Li:2001}. Nevertheless, at 3.6 and
4.5$\micron$ this extinction correction is typically around
0.005\,mag.

\subsection{Error analysis}
The pipeline also estimates the uncertainty in $\mu$, $\epsilon$ and
PA at different radii. For $\epsilon$ and PA we rely on the errors
determined by the {\sc ellipse} task, which result from the internal
errors in the harmonic fit (\citealt{Busko:1996}). As for $\mu$, the
error provided by {\sc ellipse} is derived from the rms of the pixel
values along each isophote, so it mostly reflects azimuthal variations
of the stellar emission at a given radius rather than the true
uncertainty of the mean surface brightness at that radius. We
therefore opted to measure the error in $\mu$ ourselves.

If $I$ is the incident pixel intensity, prior to sky subtraction, and
$I_{\mathrm{sky}}$ is the sky level, then the surface brightness is
  given by:
\begin{equation}
\mu = -2.5 \log(I - I_{\mathrm{sky}}) + K
\end{equation}
where $K$ includes the magnitude zero-point plus any other global
multiplying factor, such as the aperture corrections mentioned
before. Following the methodology detailed in \cite{Gil-de-Paz:2005}
and \cite{Munoz-Mateos:2009a}, the uncertainty in $\mu$ can be
estimated as:
\begin{equation}
\Delta \mu=\sqrt{(\Delta K)^2+\left(\frac{2.5\log(e)}{I-I_{\mathrm{sky}}}\right)^2 (\Delta I^2+\Delta I^2_{\mathrm{sky}})}\label{eq_delta_mu}
\end{equation}

The error in $K$ is dominated by the uncertainty in the aperture
corrections, and is of the order of 5-10\% as explained in the
previous section. Since any change in $K$ will merely translate into a
global offset of the radial profiles, the values of $\Delta \mu$ in
our ASCII tables and plots do not explicitly include the contribution
of $\Delta K$.

We computed $\Delta I$, which is the error in the incident pixel
intensity, by assuming poissonian statistics:
\begin{equation}
\Delta I = \sqrt{\frac{I}{g_{\mathrm{eff}} N_{\mathrm{isophote}}}}
\end{equation}
where $g_{\mathrm{eff}}$ is the effective gain and
$N_{\mathrm{isophote}}$ is the number of pixels within each
isophote. The effective gain, in turn, was derived from the nominal
detector gain and the exposure time of each pixel according to the weight
maps.

The term $\Delta I_{\mathrm{sky}}$ represents the uncertainty in the
background level, and was obtained as:
\begin{equation}
\Delta I_{\mathrm{sky}} = \sqrt {\frac{\sigma_{\mathrm{local}}^2}{N_{\mathrm{isophote}}}+\max \left(\sigma^2_{\mathrm{large}}-\frac{\sigma_{\mathrm{local}}^2}{N_{\mathrm{box}}},0\right)}\label{eq_sky_err}
\end{equation}
where $\sigma_{\mathrm{local}}$ and $\sigma_{\mathrm{large}}$ are the
local and large scale background rms, respectively, as described in
Section~\ref{sec_sky}. Here $N_{\mathrm{isophote}}$ is, again, the number
of pixels along a given isophote, and $N_{\mathrm{box}}$ is the number
of pixels inside the boxes used to measure the background level.

The first term in Eq.~\ref{eq_sky_err} reflects the contribution of
the local, pixel-to-pixel noise to the final error in the surface
brightness. This contribution scales down as
$1/\sqrt{N_{\mathrm{isophote}}}$, so it becomes negligible at large
radii, where the flux is averaged over a large number of pixels. It is
the second term, the large-scale noise, the one that dominates the
error budget at large radii. Note that the measured rms between the
sky values in the different boxes is partly contaminated by the local
noise, in the sense that even if there were no true large-scale
fluctuations, the rms between the sky boxes would be
$\sigma_{\mathrm{local}}/\sqrt{N_{\mathrm{box}}}$, on average. The
correction term in Eq.~\ref{eq_sky_err} accounts for this.

\subsection{Asymptotic magnitudes}\label{sec_magnitudes}
\begin{figure}
\begin{center}
\resizebox{1\hsize}{!}{\includegraphics{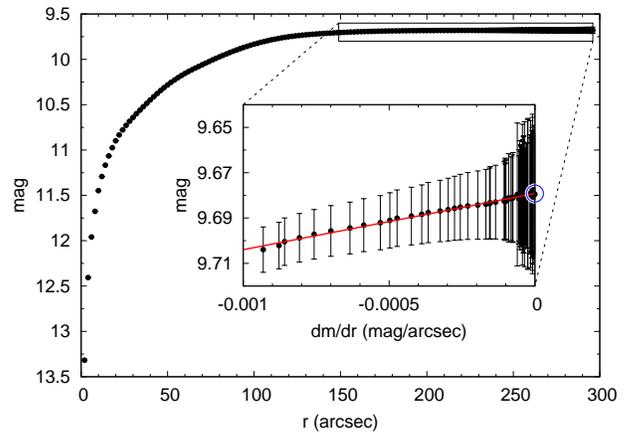}}
\caption{Illustration of how the asymptotic magnitudes are
  computed. The main panel shows the 3.6\,$\micron$ growth curve of
  NGC0936, that is, the integrated magnitude as a function of
  radius. The small inset shows the magnitude as a function of the
  magnitude gradient for the data-points in the flat portion of the
  growth curve. The red line represents a linear fit to both
  quantities. The asymptotic magnitude is the y-intercept of this fit,
  namely, the magnitude for a flat gradient (blue
  circle).\label{gcurve}}
\end{center}
\end{figure}

From the surface photometry we measure the asymptotic magnitude of
each galaxy by extrapolating the growth curve to infinity. To do this
we use the profiles with fixed $\epsilon$ and PA and 2\arcsec\ 
resolution. As an example, the bottom panel of Figure~\ref{gcurve}
shows the 3.6\,$\micron$ growth curve of NGC0936, which rises fast at
the center and then slowly approaches a flat regime. To determine this
asymptotic value, we first compute the gradient of the growth curve at
all radii. In the outer parts of the galaxy, the gradient at a given
radius and the magnitude inside that radius follow a linear trend (top
panel). We then apply a linear fit whose y-intercept (the magnitude
for a flat gradient) is by definition the asymptotic magnitude.

Note that growth curves are often used in shallow images to recover
the flux buried beneath the noise in the outskirts of galaxies. This
is not the case in the \sfg\ images, which are deep enough to always
comfortably reach the flat regime of the growth curve. Therefore, in
practice there is no extrapolation involved when fitting the growth
curves of our galaxies. In this regard, the error bugdet in the
  surface photometry is dominated by the large-scale rms, and not by
  any possible non-zero slope in the growth curve.

Once we have determined the asymptotic apparent magnitudes, we derive
the absolute ones using redshift-independent distances from NED
whenever available (79\% of the sample), and redshift-dependent ones
otherwise, assuming $H_0 = 71$\,km\,s$^{-1}$\,Mpc$^{-1}$. For those
galaxies with more than one redshift-independent distance measurement
(61\% of the whole sample), the typical rms between the different
measured distances is $\sim 15$\%, which translates into an error of
$\sim 0.3$\,mag in the absolute magnitude. While this is not strictly
speaking a true formal uncertainty, for such nearby galaxies it
provides a more realistic estimate of the actual distance error than
the uncertainty in the radial velocity, which is just $\sim 0.3$\% for
our galaxies.

\begin{figure}
\begin{center}
\resizebox{1\hsize}{!}{\includegraphics{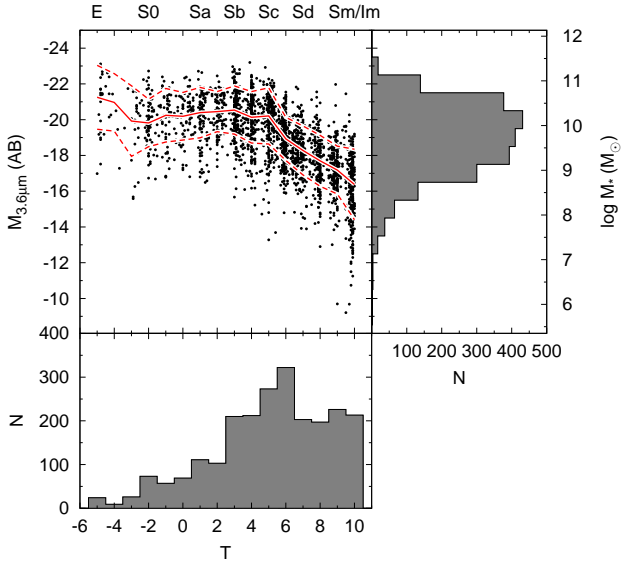}}
\caption{Stellar mass and absolute magnitude at 3.6$\micron$ of all
  \sfg\ galaxies as a function of their morphological type, together
  with the corresponding histograms of both quantities. The solid and
  dashed lines show the median trend and rms in bins of $\Delta T =
  1$.\label{morpho_mass}}
\end{center}
\end{figure}

In Fig.~\ref{morpho_mass} we plot the 3.6$\micron$ absolute magnitude
for all our galaxies as a function of their optical morphological type
as given by HyperLEDA (the 4.5$\micron$ values follow a very similar
trend). The conversion to stellar mass in the rightmost vertical axis
was done via the $M_{\star}/L$ value of \cite{Eskew:2012}. At the
late-type end of the sequence, the stellar mass rises monotonically
from $\sim 5 \times 10^8$\,$M_{\sun}$ for irregular galaxies to $1-2
\times 10^{10}$\,$M_{\sun}$ for Sc ones. The trend then remains
remarkably flat out to S0 galaxies, beyond which the stellar mass
increases again for elliptical galaxies. The rms in stellar mass for
each bin of Hubble type is typically between 0.5-0.6\,dex, or a factor
of 3-4.

\subsection{Concentration indices}
Central light concentration is one of the parameters that varies most
prominently along the Hubble sequence, and was in fact used early to
modify Hubble's original classification scheme
(\citealt{Morgan:1958}). Light concentration and other non-parametric
estimators such as the asymmetry, clumpiness, second-order moment or
Gini coefficient, have been extensively used over the years to
quantify the morphology of nearby and distant galaxies
(\citealt{Bershady:2000,Conselice:2003,Conselice:2003a,Abraham:1996,Abraham:1996a,Abraham:2003,Lotz:2004,Taylor-Mager:2007,Munoz-Mateos:2009a,Holwerda:2011}). In
particular, \cite{Holwerda:2013} presented a detailed study of the
quantitative morphology of the \sfg\ galaxies through a suite of
commonly used non-parametric estimators.

Here we present the concentration indices derived from the growth
curves measured by our pipeline. Among the different definitions
existing in the literature, we settled on $C_{31}$
(\citealt{de-Vaucouleurs:1977}) and $C_{82}$ (\citealt{Kent:1985}),
defined as follows:
\begin{eqnarray}
C_{31}&=&\frac{r_{75}}{r_{25}}\\
C_{82}&=&5 \log \left(\frac{r_{80}}{r_{20}}\right)
\end{eqnarray}
where $r_x$ is the semi-major axis of the ellipse enclosing $x$\% of
the total luminosity of the galaxy. There is some discrepancy in the
literature regarding how the total luminosity is measured when
computing concentration indices; here we simply use the asymptotic
magnitudes described before. Also, unlike concentration indices
derived from S\'{e}rsic fitting, our values are truly non-parametric,
in the sense that we do not assume any functional form for the radial
profiles or the growth curves.

For each galaxy and band we computed both concentration indices. The
results are presented in Sec.~\ref{sec_cindex}, where we investigate
why galaxies with the same stellar mass can exhibit vastly different
morphologies and concentration indices.

\subsection{Galaxy sizes and shapes}\label{size_shape}
From the growth curves we have also obtained effective radii
($r_{\mathrm{eff}}$) containing half of the total asymptotic
luminosity, independently at both IRAC bands. Radial gradients in age,
metallicity and internal extinction have a negligible impact at these
wavelengths; therefore, our half-light radii can be safely interpreted
as half-mass radii. Given that we measured $r_{\mathrm{eff}}$ along
the major axis of elliptical isophotes, our values do not need to be
corrected for inclination, as is the case in studies employing
circular apertures. Also, since we do not perform any sort of
S\'{e}rsic or similar fitting, the values that we provide constitute
the true effective radius of each galaxy, and not the effective radius
of a simple model fitted to that galaxy. In Sect.~\ref{sec_mass_size}
we discuss in detail the stellar mass-size relation of the \sfg\
galaxies for different morphologies.

The effective radii depend on the level of central concentration, in
the sense that highly-concentrated galaxies tend to have small values
of $r_{\mathrm{eff}}$ compared to the overall extent of those
galaxies. Therefore, it is also desirable to measure the global sizes
and shapes of our galaxies, including their outermost parts. Here we
follow an RC3-like approach and measure the radius, ellipticity and
position angle at 25.5 and 26.5\,\maggie, both at 3.6 and 4.5$\micron$
(that is, four sets of measurements for each galaxy). We recommend
using the 25.5 values at 3.6$\micron$ to characterize the overall
extent of the \sfg\ galaxies. Even though there is always plenty of
detectable emission beyond that surface brightness level, ellipse
fitting becomes less reliable and more sensitive to convergence
issues, object masking and background subtraction. The distribution of
isophotal sizes as a function of stellar mass and morphology is
presented in Sect.~\ref{sec_mass_size}.

Regarding the outer isophotal ellipticities, they can be used as a
photometric proxy for the disk inclination in the case of spiral
galaxies. We have in fact used the P3 ellipticities to deproject our
galaxies and recover the intrinsic shapes of rings and bars, for
instance (\citealt{Comeron:2013, Munoz-Mateos:2013}).

Note, however, that this is complicated by the fact that (a) disks may
not be perfectly circular when viewed face-on, and (b) the intrinsic
vertical thickness of disks and bulges/haloes will distort the
outermost isophotes at high inclinations (see, e.g.,
\citealt{Ryden:2004,Ryden:2006} and references therein for a
discussion on both issues). The influence of the vertical thickness is
particularly obvious in our deep IRAC images, given that thick disks
and spheroids are primarily composed of old stellar populations, which
show up prominently at these wavelengths. Users should therefore keep
this in mind when studying highly-inclined galaxies in our sample
(\citealt{Comeron:2011}). We will present inclinations derived from
multicomponent 2D fitting of the galaxy images in a forthcoming paper
(Salo et al$.$ 2014, in prep.).

\section{Results and discussion}\label{sec_results}

\subsection{Variations in concentration and morphology at fixed
  stellar mass}\label{sec_cindex}

\begin{figure*}
\begin{center}
\resizebox{1\hsize}{!}{\includegraphics{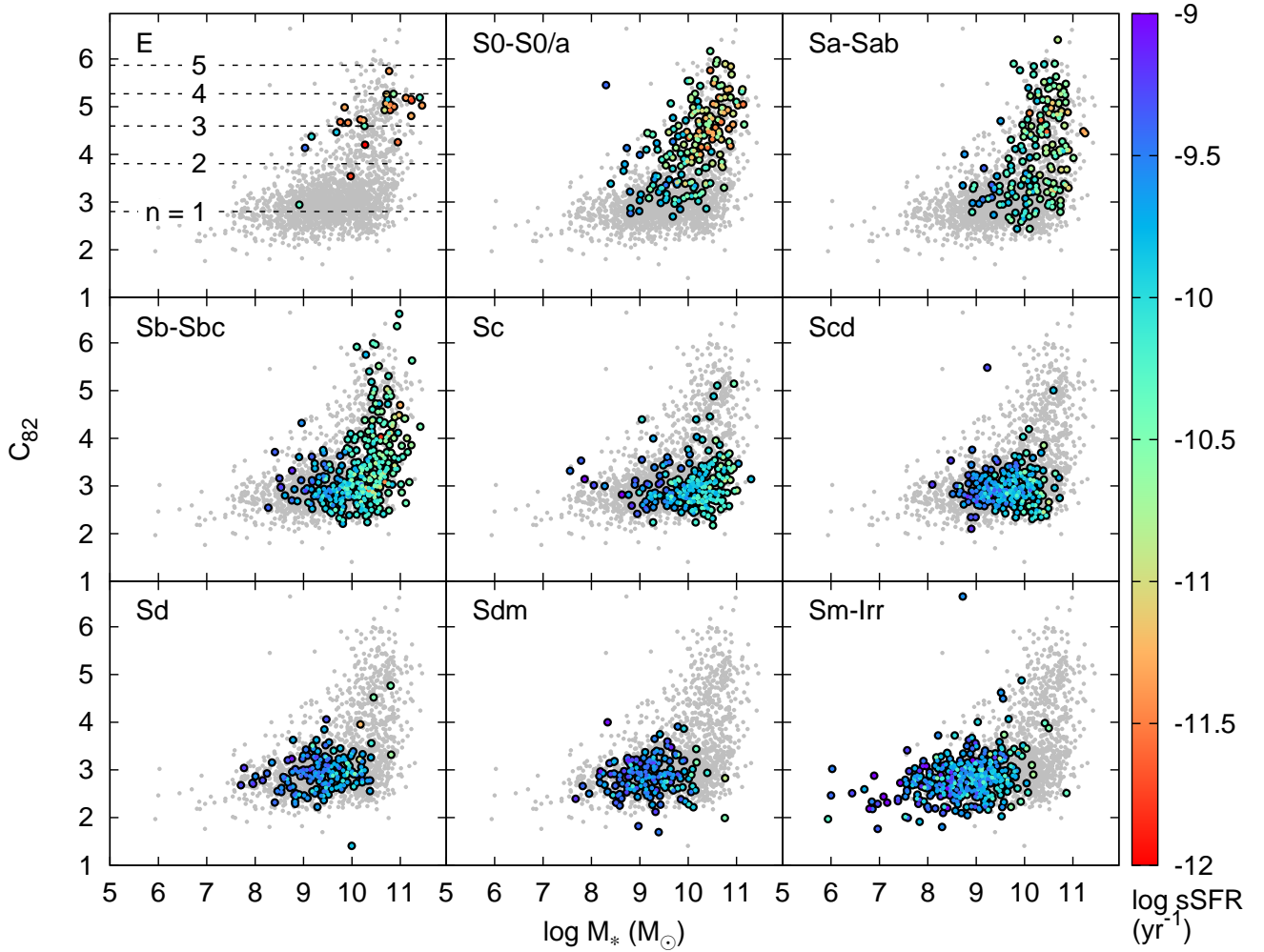}}
\caption{Concentration index $C_{82}$ at 3.6$\micron$ as a function of
  total stellar mass. The gray datapoints show the full \sfg\ sample,
  whereas the colored ones correspond to galaxies with different Hubble
  types, as indicated in each panel. The color indicates the
  extinction-corrected specific SFR, computed from GALEX and Spitzer
  data as explained in the text. As a reference, in the top-left
  panel we plot with dashed horizontal lines the theoretical
  concentration indices for S\'{e}rsic profiles with different indices
  $n$.\label{cindex_mass}}
\end{center}
\end{figure*}

\begin{figure}
\begin{center}
\resizebox{1\hsize}{!}{\includegraphics{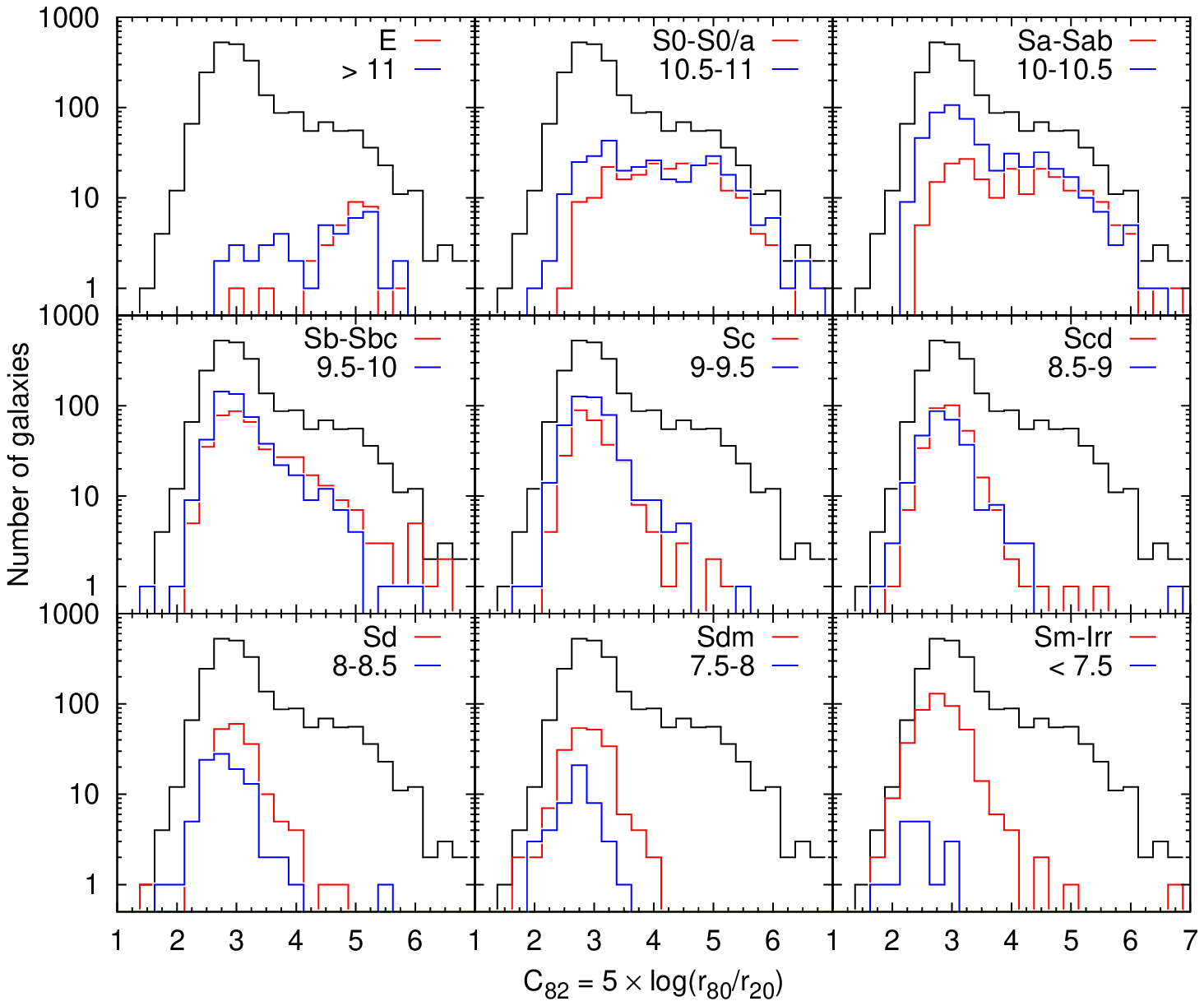}}
\caption{Histograms of the concentration index $C_{82}$ at
  3.6$\micron$ for the \sfg\ galaxies. The black histogram is the same
  in all panels, and corresponds to the full sample. The red and
  blue histograms show galaxies grouped in bins of morphological type
  and $\log M_{*}$, respectively.\label{cindex_mass_hist}}
\end{center}
\end{figure}

Figure~\ref{cindex_mass} shows the distribution of the
$C_{82}$ index at 3.6$\micron$ as a function of the stellar mass for
all \sfg\ galaxies. Figure~\ref{cindex_mass_hist} displays the same
information in histogram form. None of these plots change noticeably
when using $C_{31}$ and/or the 4.5$\micron$ data instead.

The concentration index distribution exhibits a pronounced narrow peak
centered at $C_{82} \sim 2.8$-$3$, and a much broader one peaking at
$C_{82} \sim 4.5$-$5$. The first group corresponds to disk-dominated
galaxies; indeed, the theoretical concentration index for a perfectly
exponential profile is 2.8. The second group comprises bulge-dominated
galaxies and ellipticals. The ``L''-shaped distribution of
concentration vs. stellar mass shown in Fig.~\ref{cindex_mass} nicely
describes how the spatial distribution of old stars within galaxies
varies across the Hubble sequence. For galaxies later than Sc, the
concentration index remains roughly constant at the value expected for
disk-dominated galaxies, with some scatter but no obvious dependence
on the stellar mass. But for galaxies earlier than Sc the situation
changes, and now the concentration index ranges anywhere from a
disk-dominated profile to a highly concentrated one. A similar
  distribution was found by \cite{Scodeggio:2002} in their analysis of
the $H$-band photometry of galaxies in nearby clusters. They found
that galaxies with $L_H < 10^{10} L_\sun$ exhibit a small and constant
concentration index, as well as blue optical-IR colors. At $L_H >
10^{10} L_\sun$ galaxies were found to span a large range of
concentration indices, with redder colors.

\begin{figure}
\begin{center}
\resizebox{1\hsize}{!}{\includegraphics{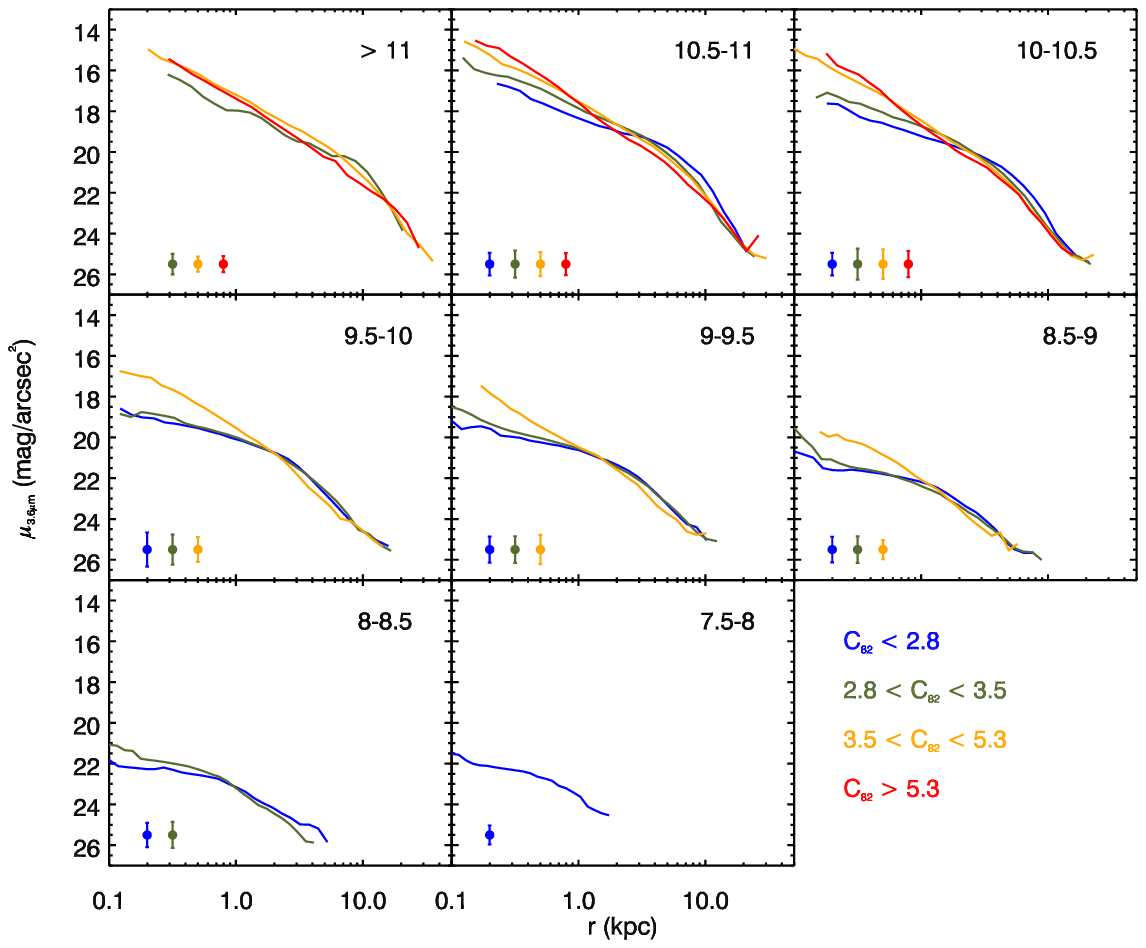}}
\caption{Median surface brightness profiles of the \sfg\ galaxies in
  bins of $\log M_{\sun}$, as indicated by the numbers at the
  top-right of each panel, and concentration index $C_{82}$, according
  to the color scheme in the legend. Prior to computing the median,
  the profiles were corrected for inclination and resampled to a
  commom physical scale in kpc. At least five galaxies were required
  in each bin to derive the median. The errorbars in each panel show
  the $\pm 1\sigma$ deviation around the median profile in each
  case.\label{median_profs}}
\end{center}
\end{figure}

To better quantify the structural variety of the \sfg\ galaxies, in
Fig.~\ref{median_profs} we show the median profiles after grouping the
galaxies in bins of stellar mass and concentration. Based on the fact
that $C_{82}=2.8$ for a pure exponential profile and $C_{82}=5.3$ for
a de Vaucouleurs one, we use these values to define four bins of mass
concentration, as shown in Fig.~\ref{median_profs}: blue and green
curves are the median profiles of disk-dominated galaxies, with the
blue ones being less concentrated than the green ones. On the other
hand, orange and red curves correspond to bulge-dominated galaxies,
with the latter being the most concentrated ones.

In galaxies more massive than $10^{10}\,M_{\sun}$, the observed
differences in concentration at a fixed mass are driven by variations
in the global radial stellar structure over scales of many kpc, all
the way from the center of the galaxies to their outer parts. However,
for galaxies less massive than $10^{10}\,M_{\sun}$ the situation is
different: differences in concentration between the blue and green
curves (both disk-dominated) are mostly due to the presence or absence
of central bright features smaller than 1\,kpc; the disks at
$r>1\,\mathrm{kpc}$ are otherwise very similar in terms of surface
density and slope.

In the following two subsections we investigate in more detail why
galaxies with the same stellar mass and morphological type present
such varied radial structure and concentration indices.

\subsubsection{Spread in concentration above $10^{10}\,M_{\sun}$}\label{sec_cindex_massive}
According to Fig.~\ref{cindex_mass}, the structural transition between
low- and high-concentration galaxies occurs at stellar masses between
$10^{10}$ and $10^{11}\,M_{\sun}$. Previous studies based on optical
imaging and spectroscopy have found abrupt transitions in the stellar
ages and star formation histories in the same mass interval
(\citealt{Kauffmann:2003, Brinchmann:2004}), with stellar populations
suddenly becoming older when going from low-mass, disk-dominated
galaxies to more massive, highly-concentrated ones.

To quantify changes in the stellar populations of the \sfg\ galaxies
we made use of GALEX measurements in the far and near
ultraviolet. Bouquin et al$.$ (2015, submitted) measured FUV and NUV
surface photometry and asymptotic magnitudes for the \sfg\ galaxies
following the same methodology described here for the Spitzer
images. Here we use their measurements to derive extinction-corrected
star formation rates (SFR).

For each galaxy we first estimated the internal dust extinction via
the integrated FUV$-$NUV color, following the prescription calibrated
by \cite{Munoz-Mateos:2009} on nearby galaxies. While this recipe is a
good proxy for internal extinction in normal star-forming spirals, it
overestimates the true extinction in elliptical galaxies, where a
significant fraction of the observed UV reddening is due to the
intrinsicially old and red stellar populations. It
also overestimates the extinction in Sdm-Irr galaxies, which tend to
have somewhat redder UV colors than other low-mass disks with similar
dust content, possibly due to differences in the extinction law, dust
geometry, or a bursty star formation activity. Therefore, we only
relied on the FUV$-$NUV color to determine the internal extinction in
the disk-dominated galaxies of our sample. Following
\cite{Munoz-Mateos:2009}, for elliptical galaxies we adopted a
constant extinction in the FUV of 2\,mags, and 0.5\,mags in the case
of Sdm-Irr galaxies. The extinction-corrected FUV luminosities were
then converted into SFR via the calibration of
\cite{Kennicutt:1998}. Note that in elliptical galaxies most of the UV
light is not associated to recent star formation, so their true SFR is
lower than the value resulting from the \cite{Kennicutt:1998}
prescription.

The data-points in Fig.~\ref{cindex_mass} are colored according to
their specific SFR (sSFR, the SFR per unit of stellar mass). The
transition from low- to high-concentration galaxies in the mass regime
between $10^{10}$ and $10^{11}\,M_{\sun}$ is accompanied by a decrease
of one order of magnitude in the sSFR, which drops from
$\sim10^{-10.5}\,\mathrm{yr}^{-1}$ in massive, low-concentration galaxies
to $\sim10^{-11.5}\,\mathrm{yr}^{-1}$ in high-concentration galaxies with
the same stellar mass. The same decrease in sSFR is found in
spectroscopic studies, where the sSFR is estimated from spectral
measurements such as the break at 4000\AA\ or the H$\alpha$
equivalent width (\citealt{Kauffmann:2003, Brinchmann:2004}).

This drop in sSFR with increasing concentration does not occur all at
once for all morphological types, though. Figure~\ref{cindex_mass}
demonstrates that it is only in S0/a galaxies and earlier types that
the sSFR drops below $\sim10^{-11}\,\mathrm{yr}^{-1}$ in
highly-concentrated objects. In these galaxies the high-concentration
and red colors result from the dominant contribution of large bulges
and stellar haloes to the total light.  But in the morphological range
between Sa and Sbc, highly-concentrated galaxies still exhibit a
relatively high sSFR of $\sim10^{-10}\,\mathrm{yr}^{-1}$, similar to
the sSFR of less concentrated galaxies in the same morphological
regime.

\begin{figure}[h]
\begin{center}
\resizebox{0.32\hsize}{!}{\includegraphics{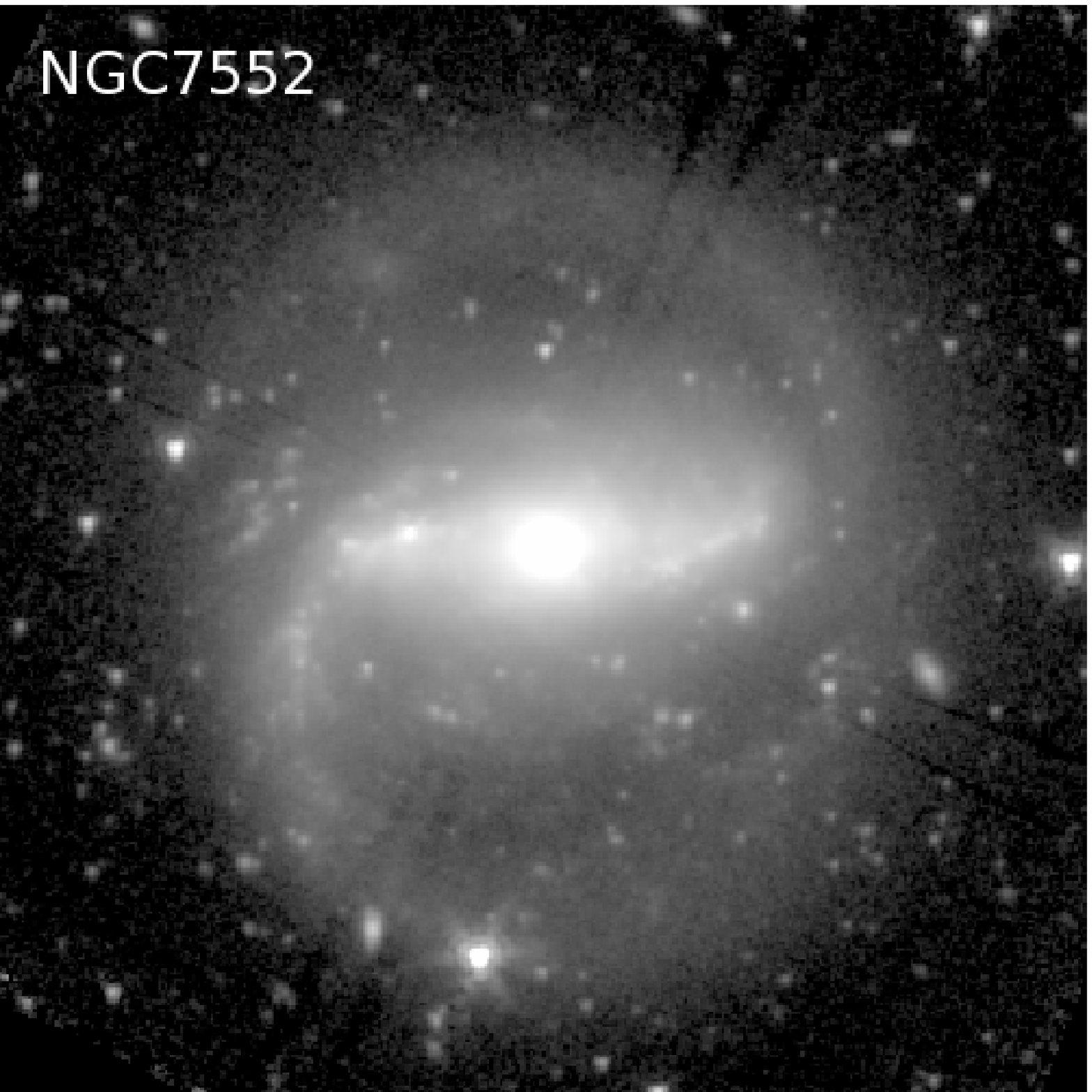}}
\resizebox{0.32\hsize}{!}{\includegraphics{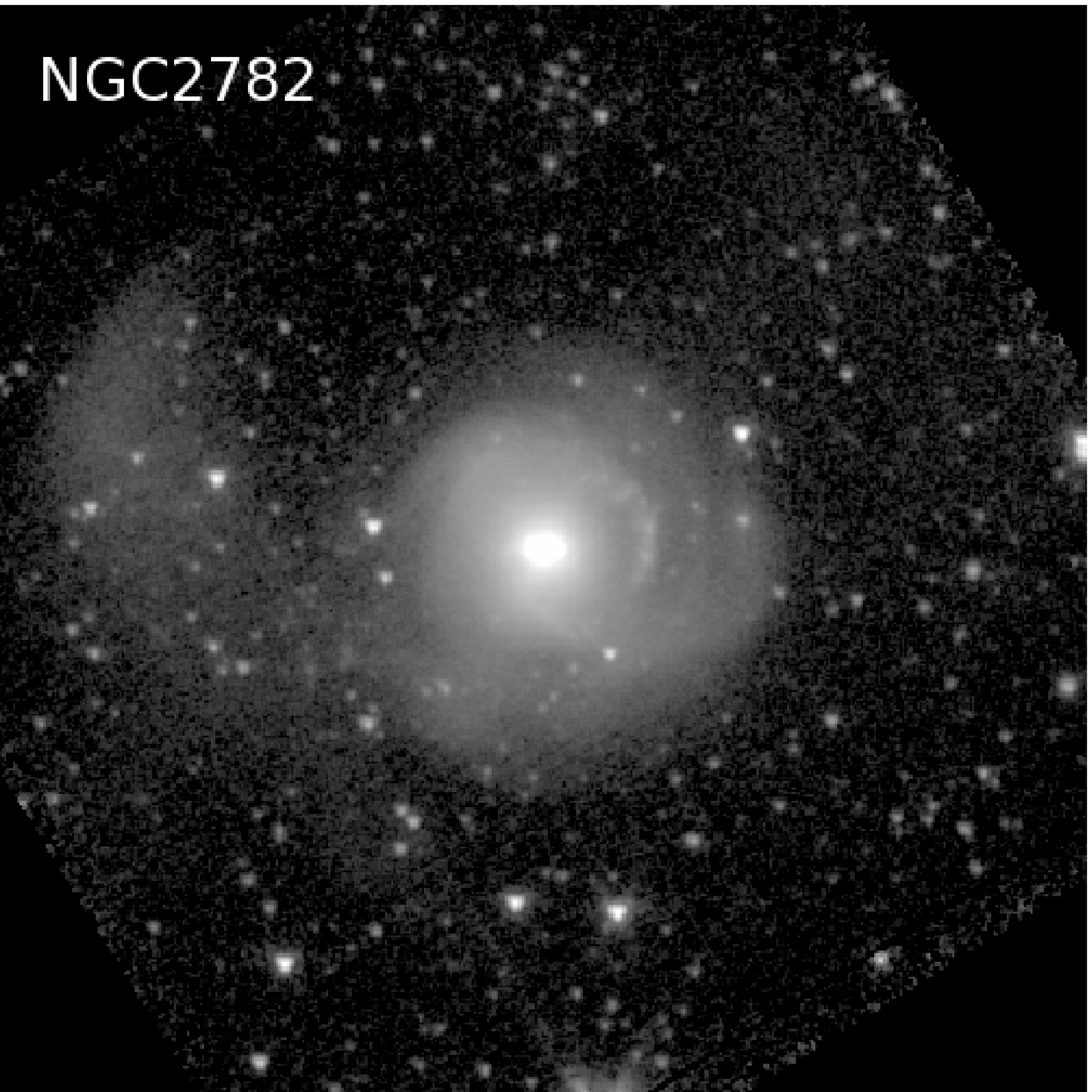}}
\resizebox{0.32\hsize}{!}{\includegraphics{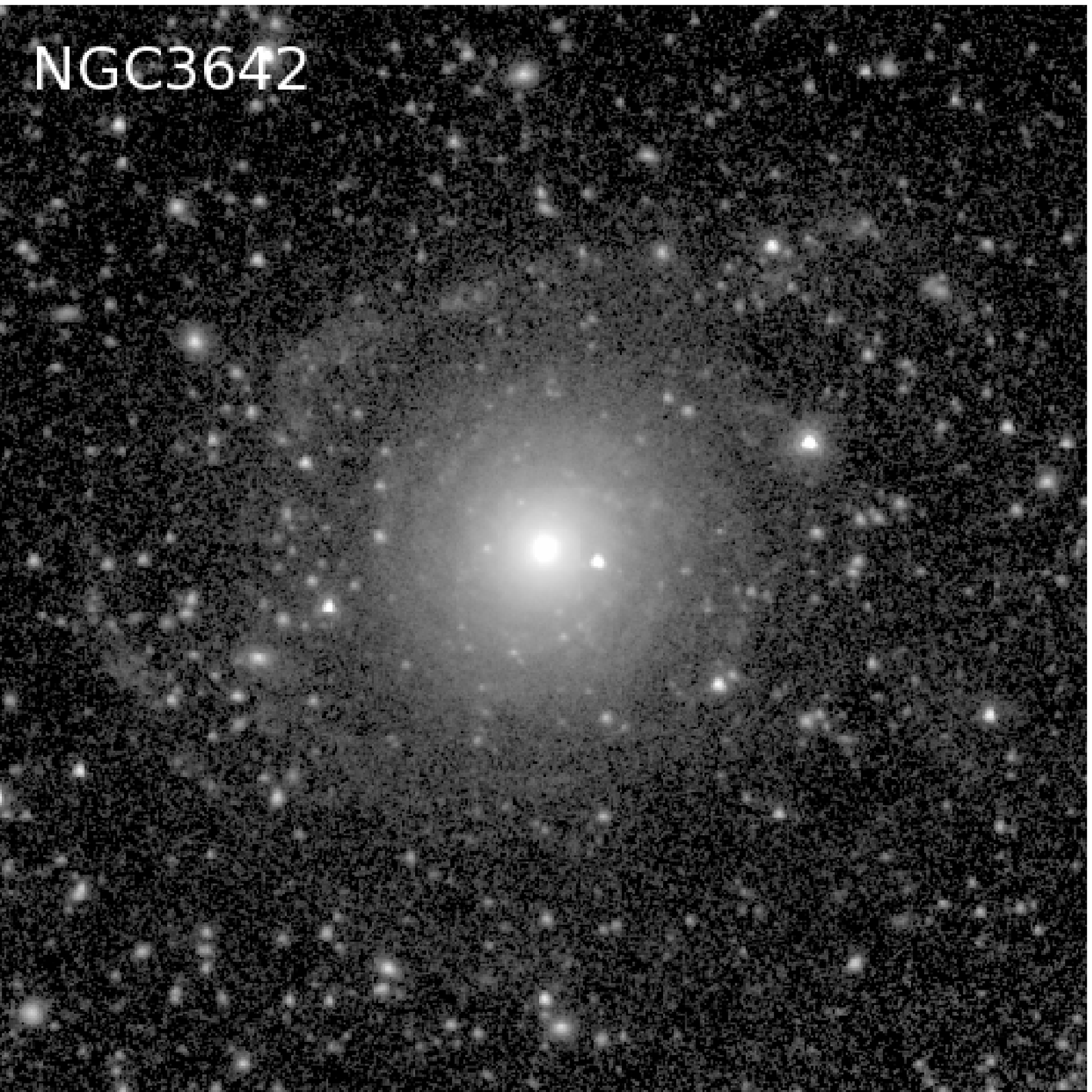}}\\
\resizebox{0.32\hsize}{!}{\includegraphics{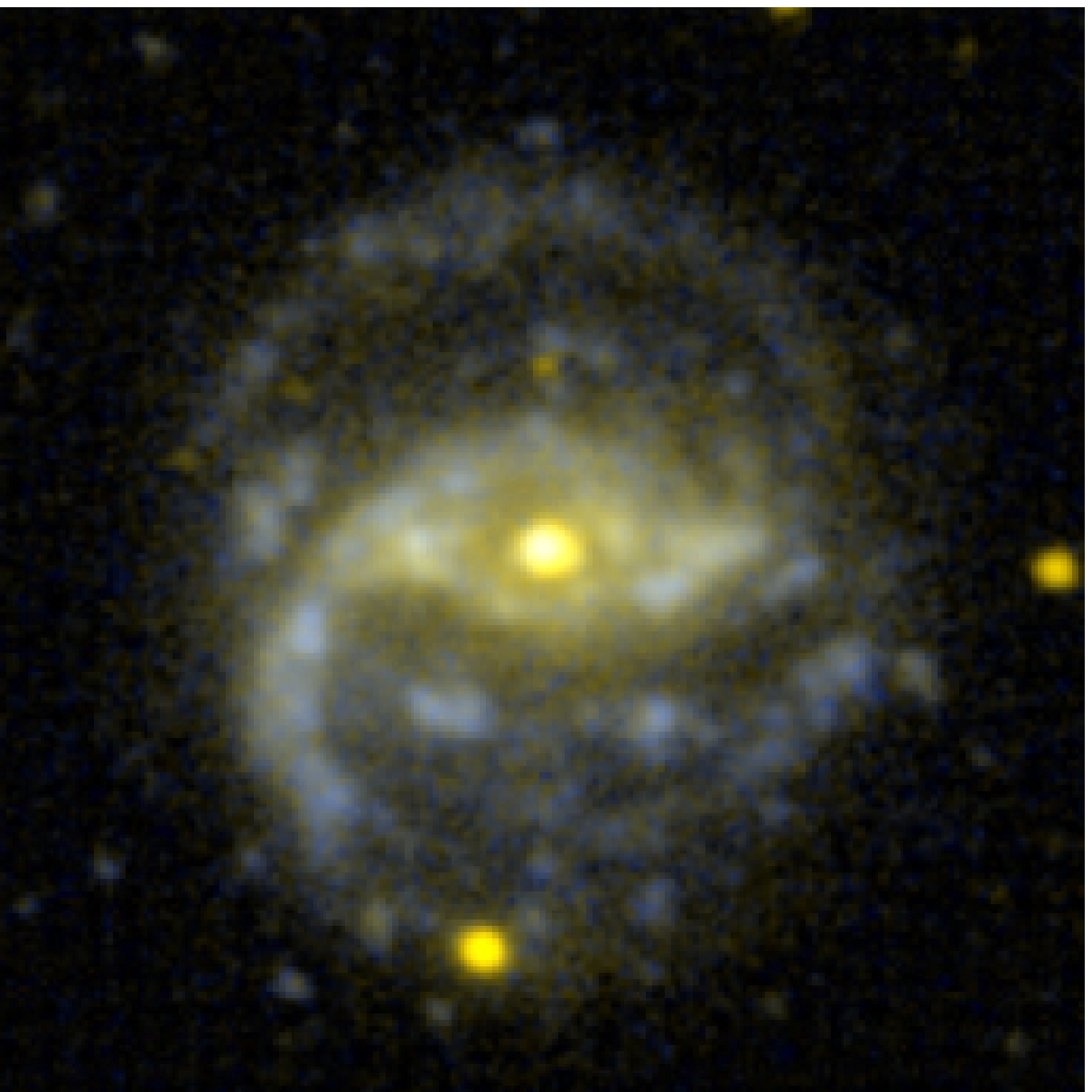}}
\resizebox{0.32\hsize}{!}{\includegraphics{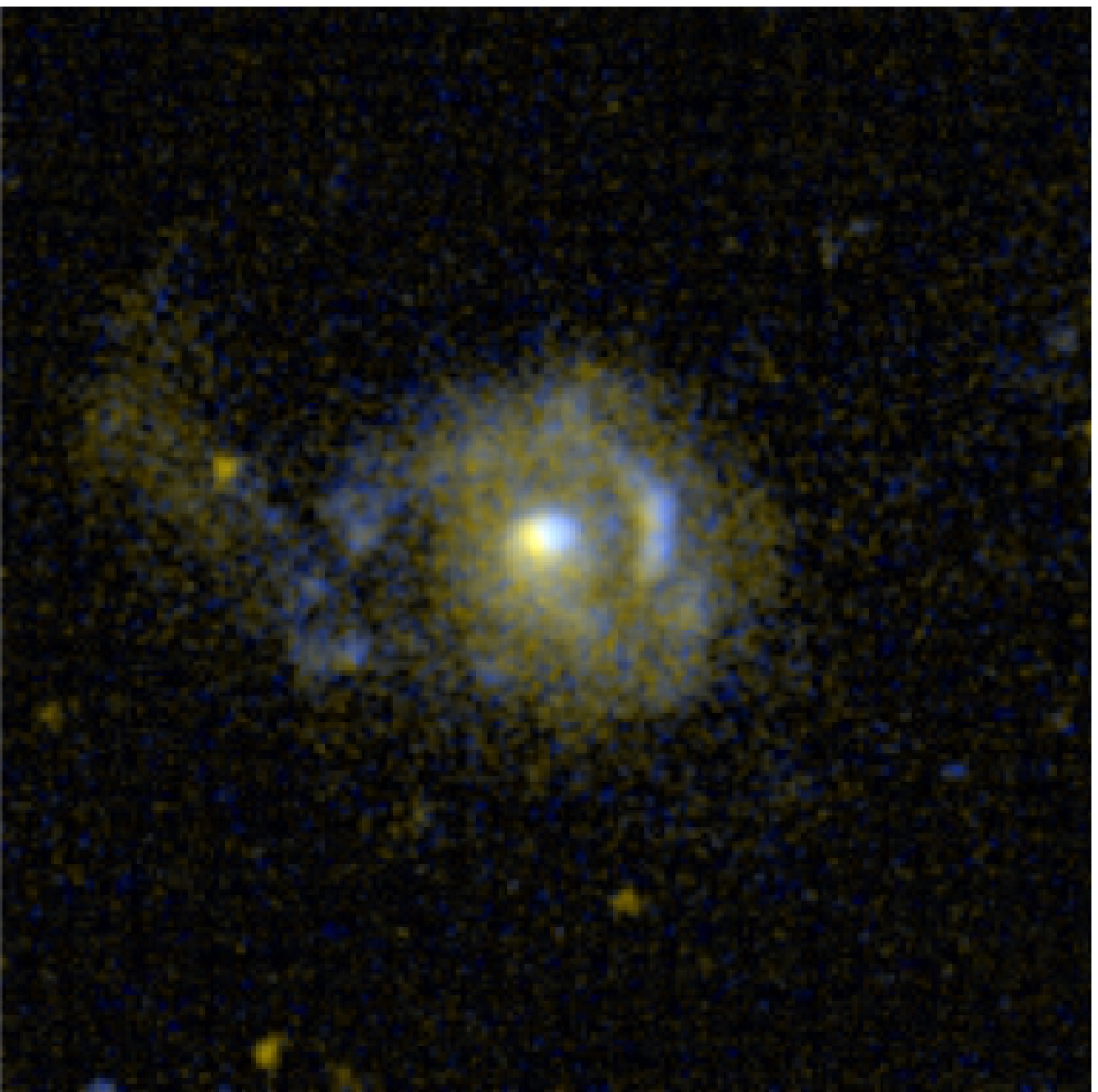}}
\resizebox{0.32\hsize}{!}{\includegraphics{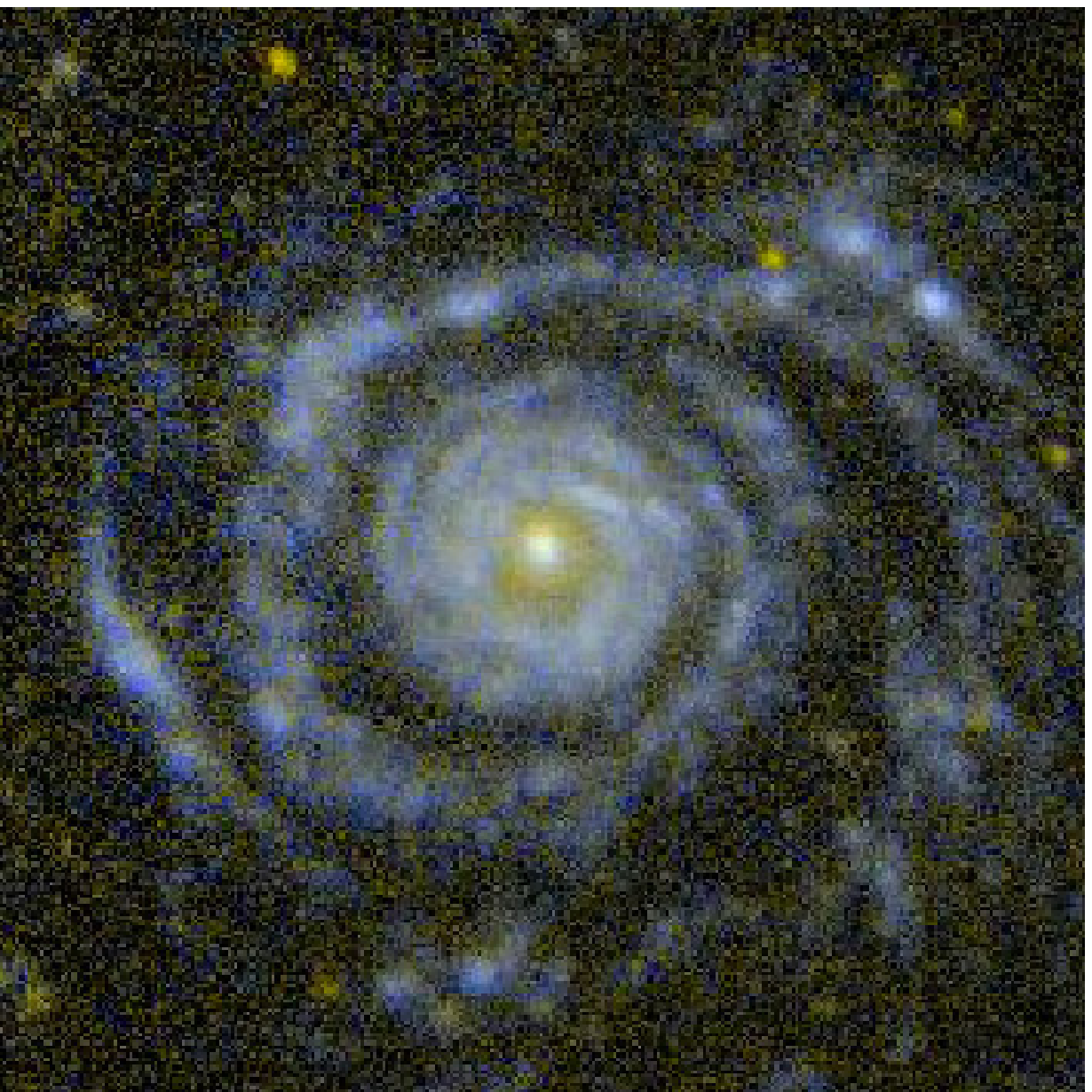}}\\
\caption{Three sample \sfg\ galaxies with high concentration indices
  at 3.6\,$\micron$ ($C_{82} \sim 5.0 - 5.5$) and also high specific
  SFR ($\lesssim10^{-10}\,\mathrm{yr}^{-1}$): NGC~7552, a barred
  spiral, NGC~2782, an interacting disk, and NGC~3642, a galaxy with a
  compact nucleus/bulge and a diffuse extended disk. The top panels
  show the \sfg\ images at 3.6\,$\micron$, whereas the bottom panels
  display false color FUV$+$NUV images from GALEX
  (\citealt{Gil-de-Paz:2007}).\label{high_cindex_sSFR}}
\end{center}
\end{figure}

What is the physical nature of these galaxies with a concentrated
stellar mass distribution but high sSFR? A visual inspection of these
objects reveals that they are a mixed bag of galaxies with very
different physical properties, which can be nevertheless sorted out
into three broad categories (Fig.~\ref{high_cindex_sSFR}):

\begin{enumerate}
\item Barred galaxies with prominent nuclear rings, such as NGC~7552,
  NGC~4593 or NGC~1365. The strong bars in these galaxies drive gas
  inwards, but this gas eventually stalls in a nuclear ring,
  most likely associated with the Inner Lindblad Resonance
  (\citealt{Regan:1999, Sheth:2000, Sheth:2005}). This
  shrinks $r_{20}$, the radius of the isophote containing 20\% of the
  total luminosity of the galaxy, thus increasing the concentration
  index $C_{82}$ in these galaxies. Note that given the intense star
  formation activity in nuclear rings and their elevated gas and dust
  content, non-stellar sources can locally dominate the 3.6 and
  4.5\,$\micron$ emission in these rings. Nevertheless, the clean
  stellar mass maps resulting from our Pipeline~5
  (\citealt{Meidt:2012, Querejeta:2014}) still reveal a prominent
  central concentration of stellar mass in these galaxies, indicating
  that bar-driven gas inflows play a key role in the secular assembly
  of (pseudo-)bulges (\citealt{Kormendy:2004, Sheth:2005}).

\item Interacting systems such as NGC~2782, NGC~7714 or
  NGC~5534. These galaxies often exhibit up-bending radial profiles,
  with a steep inner disk followed by a flatter outer one, normally
  accompanied by tidal features in the outermost regions. This
  decreases $r_{20}$ and increases $r_{80}$, yielding high
  concentration indices. Numerical simulations of minor mergers such
  as those by \cite{Younger:2007} can reproduce this radial structure:
  the interaction leads to gas inflows that contract the inner radial
  profile, while at the same time the outwards transfer of angular
  momentum expands the outer disk. Also, the sSFR is temporarily
  enhanced in these galaxies during the interaction, with prominent
  star forming knots that are clearly visible in the GALEX images.

\item Galaxies with compact bulges and smooth extended disks, often
  undisturbed as in NGC~3642. This again leads to a small $r_{20}$ and
  large $r_{80}$. While these extended outer disks often have low
  surface brightness in the infrared, they can be much brighter at UV
  wavelengths, as happens with NGC~3642. These so-called extended UV
  disks are typically found in $\sim$\,10\% - 20\% of nearby galaxies
  (\citealt{Gil-de-Paz:2005a, Thilker:2005, Thilker:2007}), but the
  origin of these galaxy-wide star formation events is still under
  debate.
 \end{enumerate}

In brief, for stellar masses above $10^{10}\,M_{\sun}$, galaxies with
a centrally concentrated radial distribution of old stars are either
(a): quiescent galaxies with a prominent stellar bulge/halo, or (b)
star-forming galaxies with central mass concentrations due to a
variety of mechanisms such as bar- or merger-driven inflows.

\subsubsection{Spread in concentration below $10^{10}\,M_{\sun}$}
Figure~\ref{cindex_mass} shows that galaxies with stellar masses below
$10^{10}\,M_{\sun}$ still exhibit a considerable scatter in the
concentration index at any given mass. While all these galaxies have
concentration indices broadly consistent with a disk-dominated
profile, the actual values can range anywhere from $C_{82} \sim 2$ to
$\sim 3.5$. Moreover, this scatter in $C_{82}$ does not seem to be
correlated with changes in the sSFR, as was the case in the more
massive galaxies analyzed before.

\begin{figure}
\begin{center}
\resizebox{1\hsize}{!}{\includegraphics{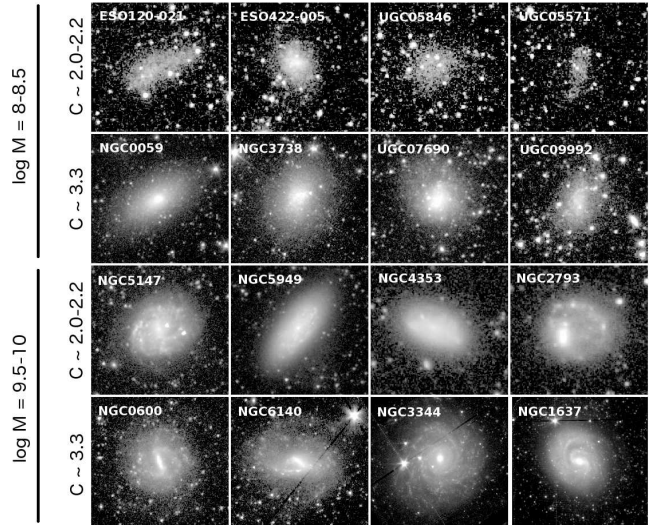}}
\caption{{\it First row:} galaxies with stellar masses between $10^8 -
  10^{8.5}\,\mathrm{M_{\sun}}$ and low concentration indices ($C_{82}
  \sim 2.0 - 2.2$). {\it Second row:} galaxies with the same stellar mass
  as those in the first row, but higher concentration values ($C_{82}
  \sim 3.3$). {\it Third row:} more massive disks with masses around $10^{9.5} -
  10^{10}\,\mathrm{M_{\sun}}$, but low concentration. {\it Fourth row:} same
  stellar mass as in the third row, but higher concentration.\label{sample_cindex}}
\end{center}
\end{figure}

To ascertain the origin of these structural variations in low mass
disks, we visually inspected galaxies with the same stellar mass but
extreme values of concentration. Figure~\ref{sample_cindex} displays
several representative examples.

The top two rows show very low-mass dwarf disks, all of them with
stellar masses between $10^8 - 10^{8.5}\,\mathrm{M_{\sun}}$. The
objects in the first row are characterized by very low concentration
indices of $C_{82} \sim 2.0 - 2.2$ (smaller than for an exponential
profile), whereas those in the second row all have $C_{82} \sim 3.3$
(larger than for an exponential). This higher concentration is due to
central aggregations of stars (usually resolved in our images) that are
absent in the first group of galaxies. GALEX images of some of these
concentrated dwarfs, like NGC~0059 and NGC~3738, reveal bluer UV
colors in the central parts than in the outskirts, possibly hinting to
an outside-in formation for these objects.

The third and fourth rows in Fig.~\ref{sample_cindex} correspond to
more massive disks, with stellar masses between $10^{9.5} -
10^{10}\,M_{\sun}$. Galaxies in the third row present $C_{82} \sim 2.0
- 2.2$, while those in the fourth row have $C_{82} \sim 3.3$. In this
case the higher concentration results from bright bars as in NGC~0600
or NGC~6140. Unbarred galaxies can be also concentrated if they host
bright and compact (pseudo-)bulges, as in NGC~3344. In contrast, low
concentrated disks in this mass range either have weakly-defined bars
(NGC~5147) or no bars at all (NGC~5949). Moreover, some galaxies such
as NGC~4353 have bright inner disks that lead to central plateaus in
their radial profiles, leading to concentration indices lower than
those expected for an exponential profile.

We also find a wealth of asymmetric galaxies which may have recently
undergone gravitational interactions. In some of them, like NGC~2793,
the stellar mass is so dislodged that the radial surface density is
almost flat, leading to very low concentration indices. On the other
hand, galaxies like NGC~1637 exhibit a much milder asymmetry, and
therefore have higher concentration values (\citealt{Zaritsky:2013}
and references therein).

In summary, even though galaxies with stellar masses below
$10^{10}\,M_{\sun}$ tend to have low concentration, at any given mass
variations in concentration are still present, due to the presence of
central star clusters, bars, pseudo-bulges, or lack thereof.

\subsection{The stellar mass-size relation}\label{sec_mass_size}
The mass-size relation provides very stringent constraints on models
of galaxy formation and evolution. From a cosmological point of view,
the trend between increasing galaxy size and mass reflects, to first
order, the physical connection between the mass of the dark matter
halo and its angular momentum. Indeed, in a $\Lambda$CDM cosmology the
characteristic size of a disk should scale as
$M^{1/3}_{\mathrm{halo}}$ (\citealt{Fall:1980, Mo:1998}). In
practice, though, the stellar mass-size relation also depends on
whether the halo gas conserves its angular momentum as it settles onto
the disk, how efficiently that gas is later converted into stars, the
impact of AGN and stellar feedback, and whether internal and external
processes rearrange stars within the galaxy.

The mass-size (or luminosity-size) relation has been extensively
explored both in the local universe (see, e.g$.$, \citealt{Kauffmann:2003, Shen:2003,
  Courteau:2007, Fernandez-Lorenzo:2013, Lange:2014}) and at high redshift
(e.g$.$ \citealt{Barden:2005, Trujillo:2004, Trujillo:2006, Franx:2008,
  van-der-Wel:2014}). Here we take advantage of the little sensitivity
of the IRAC bands to dust extinction and $M/L$ variations to derive a
self-consistent stellar mass-size relation in the local universe.

\begin{figure}
\begin{center}
\resizebox{1\hsize}{!}{\includegraphics{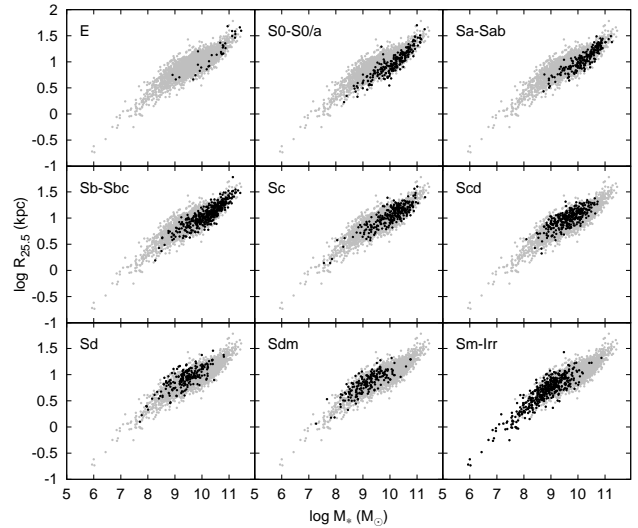}}
\caption{Semi-major radius at $\mu_{3.6} = 25.5$\,\maggie\ as a
  function of stellar mass of all \sfg\ galaxies (gray dots), grouped by
  their morphological type (black dots).\label{iso_size_mass}}
\end{center}
\end{figure}

In Fig.~\ref{iso_size_mass} we plot $R_{25.5}$, the semi-major radius
at $\mu_{3.6} = 25.5$\,\maggie, as a function of the total stellar
mass for all \sfg\ galaxies. We find an obvious and expected monotonic
trend, in the sense that galaxies with higher stellar mass are also
larger. However, the slope of this trend is not constant, as galaxies
of different Hubble types populate different areas in this plot,
defining clearly distinct sequences. As in Figs.~\ref{morpho_mass} and
\ref{cindex_mass}, Sc galaxies constitute a well-defined transition
type in this diagram, in the sense that earlier-type galaxies are up
to a factor of $\sim 10$ more massive than later-type ones for the
same outermost size $R_{25.5}$.

\begin{figure*}
\begin{center}
\resizebox{1\hsize}{!}{\includegraphics{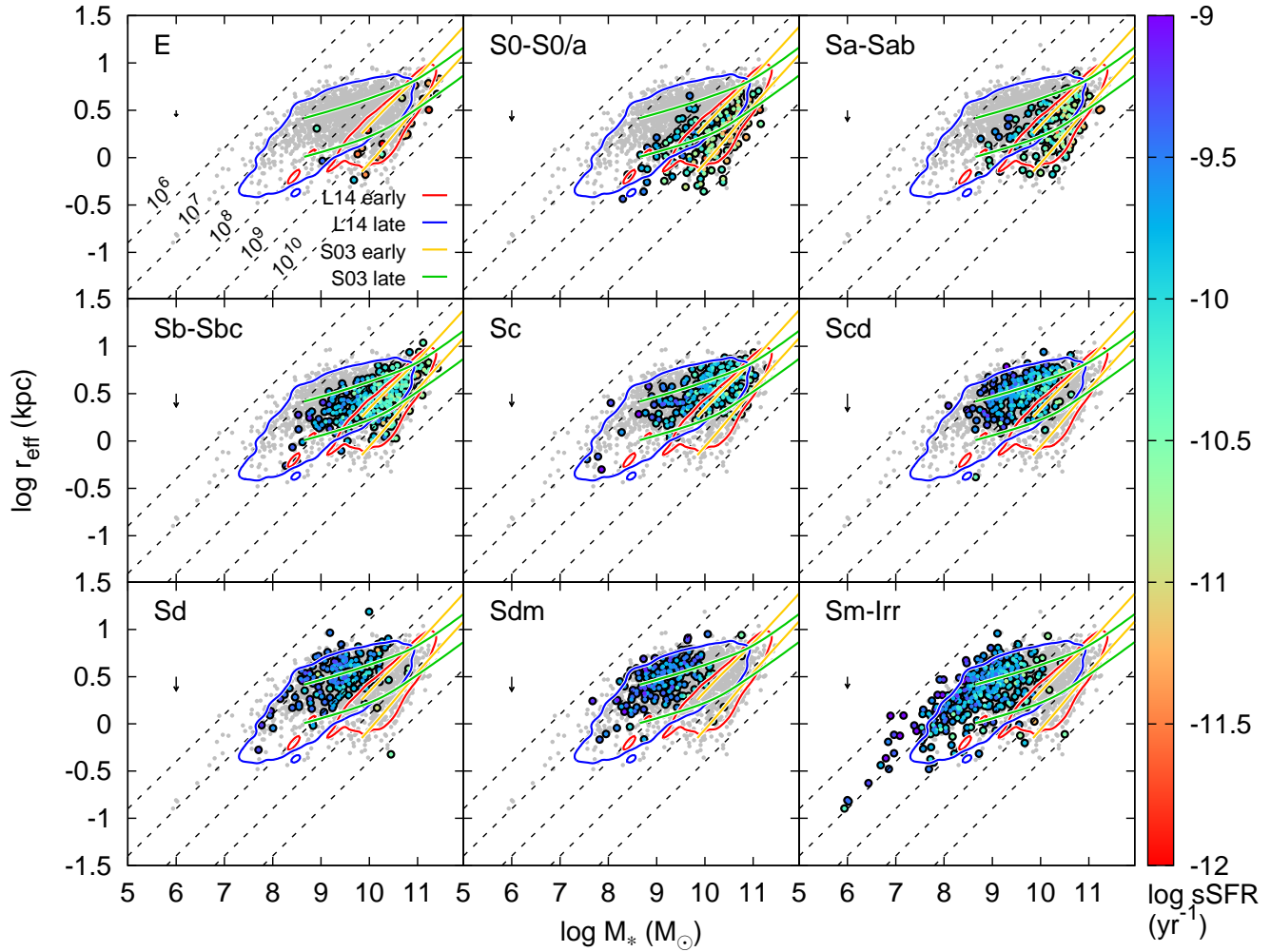}}
\caption{Effective radius along the semi-major axis at 3.6\,$\micron$
  as a function of the total stellar mass, grouped by morphological
  type. Data-points are color-coded based on the sSFR of each galaxy
  (see Sect.~\ref{sec_cindex_massive} for details). The dashed
  diagonal lines mark constant values of the average stellar mass
  surface density inside $r_{\mathrm{eff}}$. The red and blue contours
  delineate the mass-size relation of early- and late-type galaxies
  found by \cite{Lange:2014} in the GAMA survey. The orange and green
  curves correspond to the early- and late-type distributions found by
  \cite{Shen:2003} on SDSS data. The latter were derived from circular
  apertures; the vertical arrows in each panel show how our
  ellipse-based measurements would shift downwards had we used
  circular apertures, given the median ellipticity in each bin of
  morphological type.\label{reff_mass}}
\end{center}
\end{figure*}

Isophotal radii represent a robust metric of global galaxy sizes in
the nearby universe, but their applicability to distant galaxies is
hampered by cosmological surface brightness dimming. In order to
facilitate the comparison of the \sfg\ mass-size relation with
observations at higher redshits, we have also measured the effective
radius $r_{\mathrm{eff}}$ of our galaxies. In Fig.~\ref{reff_mass} we
show the trend between the effective radius at 3.6\,$\micron$ and the
total stellar mass of our galaxies, in bins of morphological type. As
explained in Sect.~\ref{size_shape}, $r_{\mathrm{eff}}$ was measured
along the semi-major axis of elliptical apertures, so no inclination
corrections are required. The diagonal lines in this plot represent different
values of the average stellar mass surface density inside
$r_{\mathrm{eff}}$, that is, $0.5 M_{\star}/(\pi
r_{\mathrm{eff}}^2)$. Points are again color-coded according to their
extinction-corrected sSFR.

The main difference with respect to the mass-size relation based on
isophotal radii (Fig.~\ref{iso_size_mass}) is an increased spread in
$M_{\star}$ at fixed size, because a higher concentration index will
significantly shrink $r_{\mathrm{eff}}$, but not so much the outer
isophotal radius. As a result, early-type galaxies can be up to two
orders of magnitude more massive than late-type ones with the same
$r_{\mathrm{eff}}$.

As a comparison, we have overplotted the mass-size relations for
early- and late-type galaxies found by \cite{Lange:2014} at $0.01 < z
< 0.1$, based on data from the Galaxy and Mass Assembly survey (GAMA,
\citealt{Driver:2011}). These authors relied on different criteria to
separate early-type galaxies from late-type ones: S\'{e}rsic index,
colors and visual identification; here we plot the latter. We show
their mass-size relation in the $r$-band (where their sample selection
was performed), but shifting $r_{\mathrm{eff}}$ down by 0.076\,dex to
account for the slightly smaller sizes of galaxies at
3.6\,$\micron$. This offset was derived from the empirical fits of
$r_{\mathrm{eff}}$ vs$.$ wavelength in \cite{Lange:2014}.

We find an excellent agreement between the \sfg\ mass-size relation
and the GAMA one. Massive early-type galaxies exhibit a roughly
constant average stellar surface density of
$10^9\,M_{\sun}\,\mathrm{kpc}^{-2}$, whereas in late-type galaxies the
surface density decreases monotonically from
$10^9\,M_{\sun}\,\mathrm{kpc}^{-2}$ in massive disks to
$10^7\,M_{\sun}\,\mathrm{kpc}^{-2}$ in low-mass disks
(\citealt{Kauffmann:2003}). The transition from the early- to the
late-type sequence occurs at a specific SFR of $\sim
10^{-11}\,\mathrm{yr}^{-1}$. In agreement with our findings when
discussing the spread in concentration at fixed stellar mass, here we
also note that the early-type mass-size sequence is not entirely
populated by red and old systems: bars, mergers and XUV-disks can lead
to relatively high sSFR in galaxies that have otherwise high
infrared concentration indices and stellar mass surface densities (see
Fig.~\ref{high_cindex_sSFR}).

A similar mass-size relation was previously found by \cite{Shen:2003},
using SDSS data of nearby galaxies as well. In Fig.~\ref{reff_mass} we
have overplotted their trends ($\pm 1\sigma$) for early- and late-type
galaxies, which were classified as such based on their concentration
and S\'{e}rsic indices. It is worth noting that \cite{Shen:2003} used
circular apertures instead of elliptical ones, an this typically
decreases the effective radius as $r_\mathrm{eff\, circ} =
r_\mathrm{eff} \sqrt{b/a}$, where $b/a$ is the axial ratio. We
computed the median offset that our data-points would undergo had we
used circular apertures, based on the distribution of axial ratios in
each bin of morphological type. These offsets are shown as vertical
arrows in Fig.~\ref{reff_mass} . Offsets are negligible in early-type,
spheroid-dominated galaxies, which tend to be round regardless of the
observing angle. However, inclination does play a role in late-type,
disk-dominated galaxies, where $r_\mathrm{eff}$ can decrease by $\sim
0.1-0.2$\,dex when circular apertures are used.

\subsubsection{Direct size comparisons with previous studies}
Galaxy size measurements, and effective radii in particular, can be
biased by the image depth, the observed wavelength (due to radial
color gradients), and the fitting methodology. In order to assess the
impact of these factors in determining accurate galaxy sizes, here we
perform a direct galaxy-to-galaxy comparison of the \sfg\ effective
radii with published measurements from other surveys that overlap with
ours.

We first cross-matched our sample with the 2MASS Extended Source
Catalog (XSC; \citealt{Jarrett:2000}), which yielded 1687 galaxies in
common (unmatched galaxies typically have $m_{3.6} > 13 - 14$\,AB
mag). The XSC provides surface photometry carried out in essentially
the same way as in \sfg: first, 1D radial profiles are measured using
concentric elliptical apertures; second, total magnitudes are
obtained by fitting the outer part of the growth curve; finally,
$r_{\mathrm{eff}}$ is derived by integrating on the growth curve
until reaching the half-light value. The XSC quotes effective radii
along the major axis in $J$, $H$ and $K_S$, but the values differ by
less than $\sim 5-10$\% among the three bands, so for each galaxy we
simply averaged the three values.

\begin{figure}
\begin{center}
\resizebox{1\hsize}{!}{\includegraphics{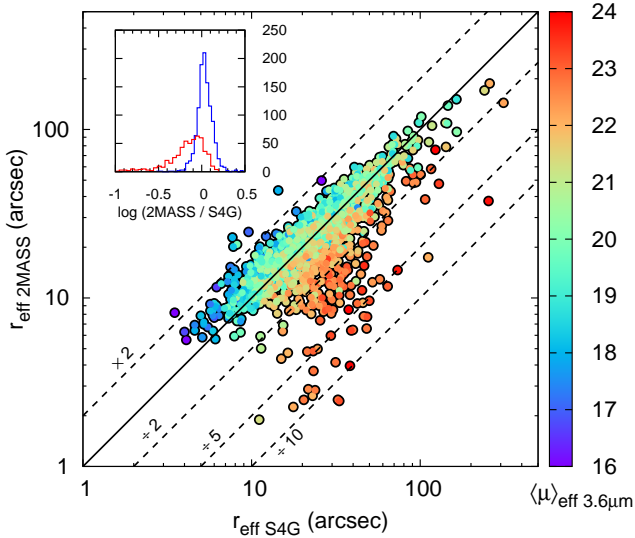}}
\caption{Comparison between the 2MASS XSC effective radii and the
  \sfg\ ones. For each galaxy, the 2MASS radius is the average of the
  effective radii in $J$, $H$ and $K_S$. The \sfg\ radii are measured
  at 3.6\,$\micron$. The color code reflects the average surface brightness
  inside the effective radius at 3.6\,$\micron$, without correcting
  for inclination, in \maggie. The inset plot shows the logaritmic
  ratio of the 2MASS and \sfg\ effective radii for galaxies with
  $\langle\mu\rangle_{\mathrm{eff}\,3.6} < 21$ (blue histogram) and $>21$
  (red histogram).\label{2MASS}}
\end{center}
\end{figure}

\begin{figure}
\begin{center}
\resizebox{0.3\hsize}{!}{\includegraphics{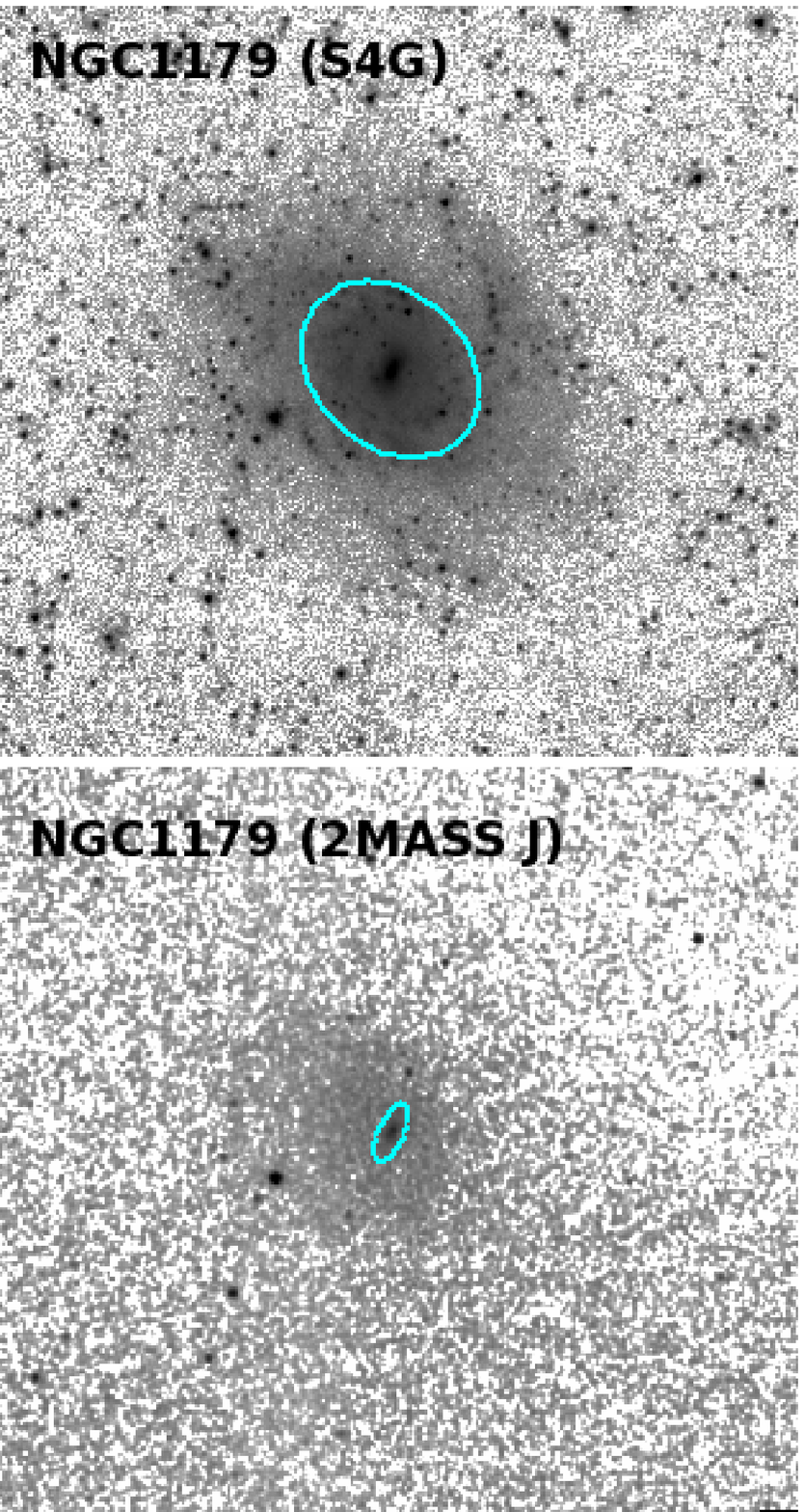}}
\resizebox{0.3\hsize}{!}{\includegraphics{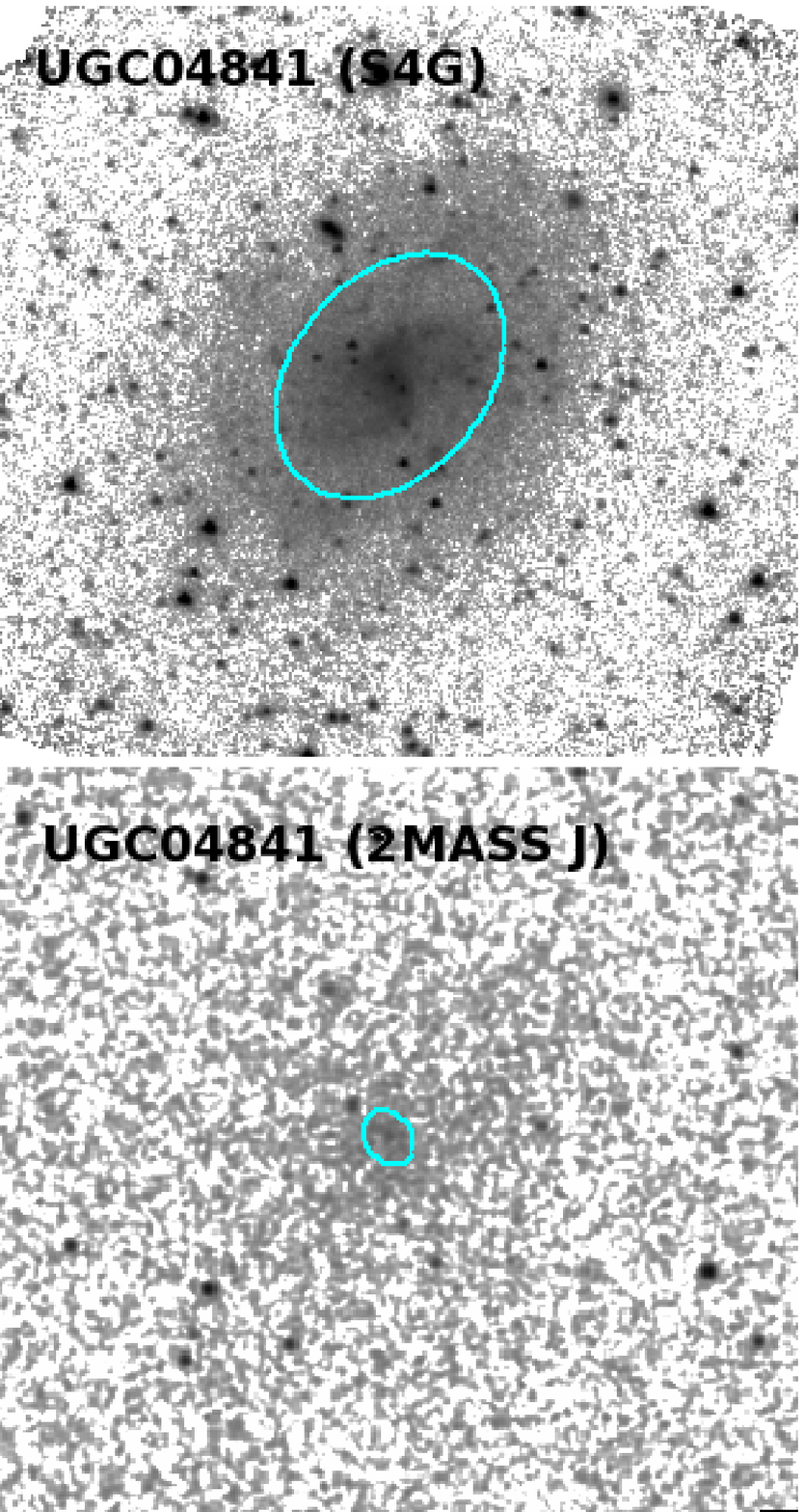}}
\resizebox{0.3\hsize}{!}{\includegraphics{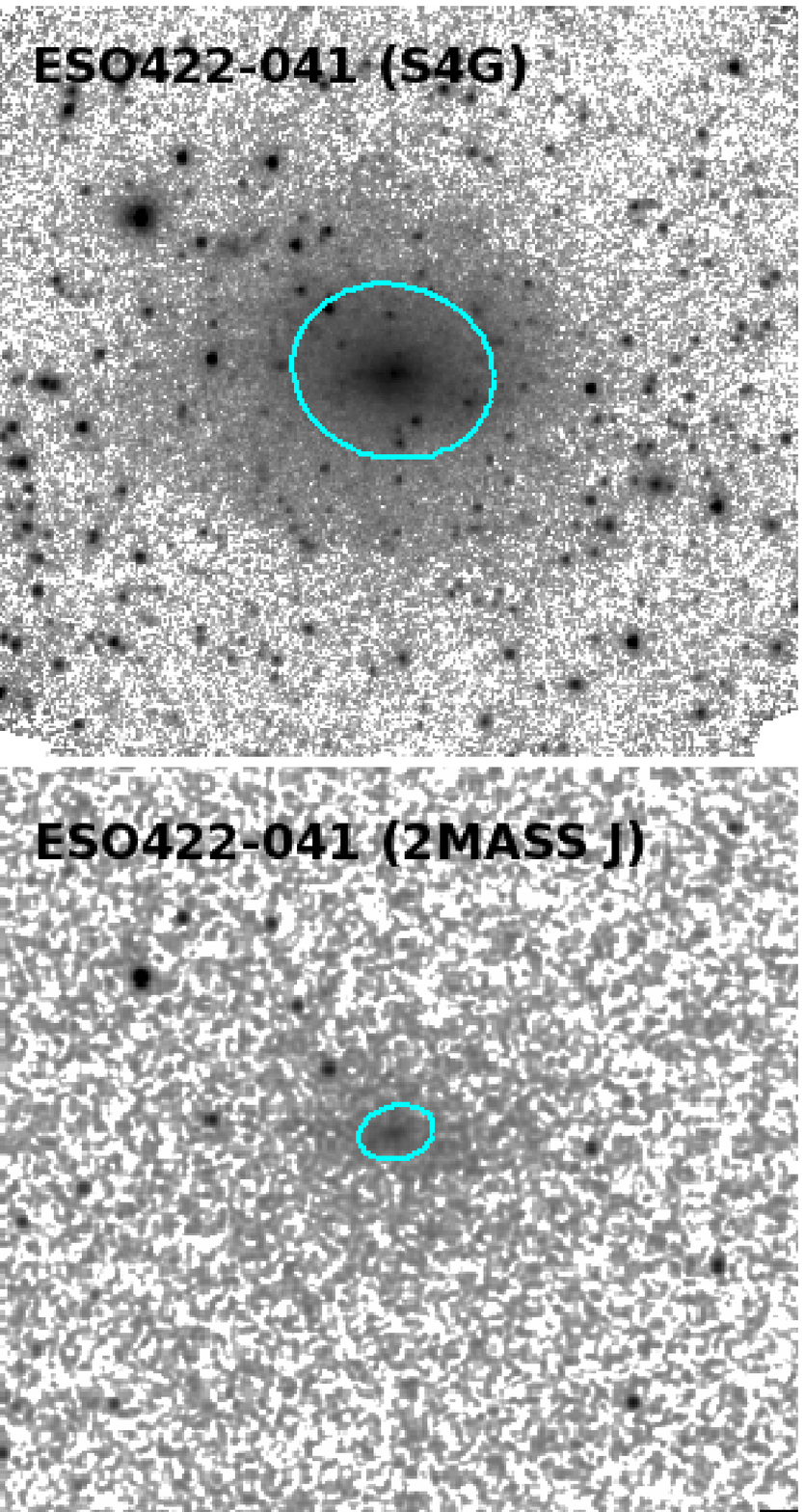}}\\
\caption{Comparison of \sfg\ 3.6\,$\micron$ images (top) and 2MASS $J$
  ones (bottom) for some representative galaxies with
  $\langle\mu\rangle_{\mathrm{eff}\,3.6} > 21$\,\maggie. Images of the
  same galaxies are displayed at the same spatial scale. The cyan
  ellipses show the respective half-light
  apertures.\label{2MASS_sample}}
\end{center}
\end{figure}

In Fig.~\ref{2MASS} we compare the \sfg\ and XSC effective
radii. Given the similar methodologies and wavelengths of both surveys
one would expect an excellent agreement between both sets of
measurements. While this is generally the case, the XSC sizes are
considerably affected by the shallowness of the 2MASS images. To
illustrate this, we have color-coded the data-points according to
$\langle\mu\rangle_{\mathrm{eff}\,3.6}$, the average surface
brightness inside the half-light elliptical aperture at
3.6\,$\micron$, uncorrected for inclination. For bright galaxies with
an effective surface brightness $\langle\mu\rangle_{\mathrm{eff}\,3.6}
< 21$\,\maggie\ the 2MASS radii agree with the \sfg\ ones with a
1$\sigma$ scatter of $\pm 20$\%. However, galaxies with fainter
surface brightness levels appear to be several times smaller in
2MASS. This is further highlighted in Fig.~\ref{2MASS_sample}, where
we compare the \sfg\ and 2MASS images of low surface brightness
galaxies whose effective radius is severely underestimated in
2MASS. This effect must be also taken into account at high redshift,
where cosmological surface brightness dimming hampers the detection of
galaxy outskirts.

We can also compare our size measurements with those obtained from
SDSS data using parametric fits, in order to understand how many
components these fits must have in order to accurately recover the
global non-parametric sizes of galaxies. The SDSS pipeline performs
simple exponential or de Vaucouleurs fits to the light profiles of
galaxies, but as we will show below these basic fits provide only
rough (and potentially biased) estimates of galaxy sizes. More
sophisticated fits, including S\'{e}rsic profiles and/or
multi-component fits, have been widely used by other authors on SDSS
data (see, e.g., \citealt{Blanton:2005, Gadotti:2009, Simard:2011,
  Kelvin:2012, Lackner:2012}).

Here we compare our effective radii with those measured by
\cite{Lackner:2012} on a sample of SDSS nearby galaxies, which
includes 124 \sfg\ objects. These authors fitted the $r$-band image of
each galaxy with five different 2D models: a single exponential model,
a single de Vaucouleurs one, a S\'{e}rsic model, an exponential bulge
plus an exponential disk, and a de Vaucouleurs bulge plus an
exponential disk.

\begin{figure*}
\begin{center}
\resizebox{1\hsize}{!}{\includegraphics{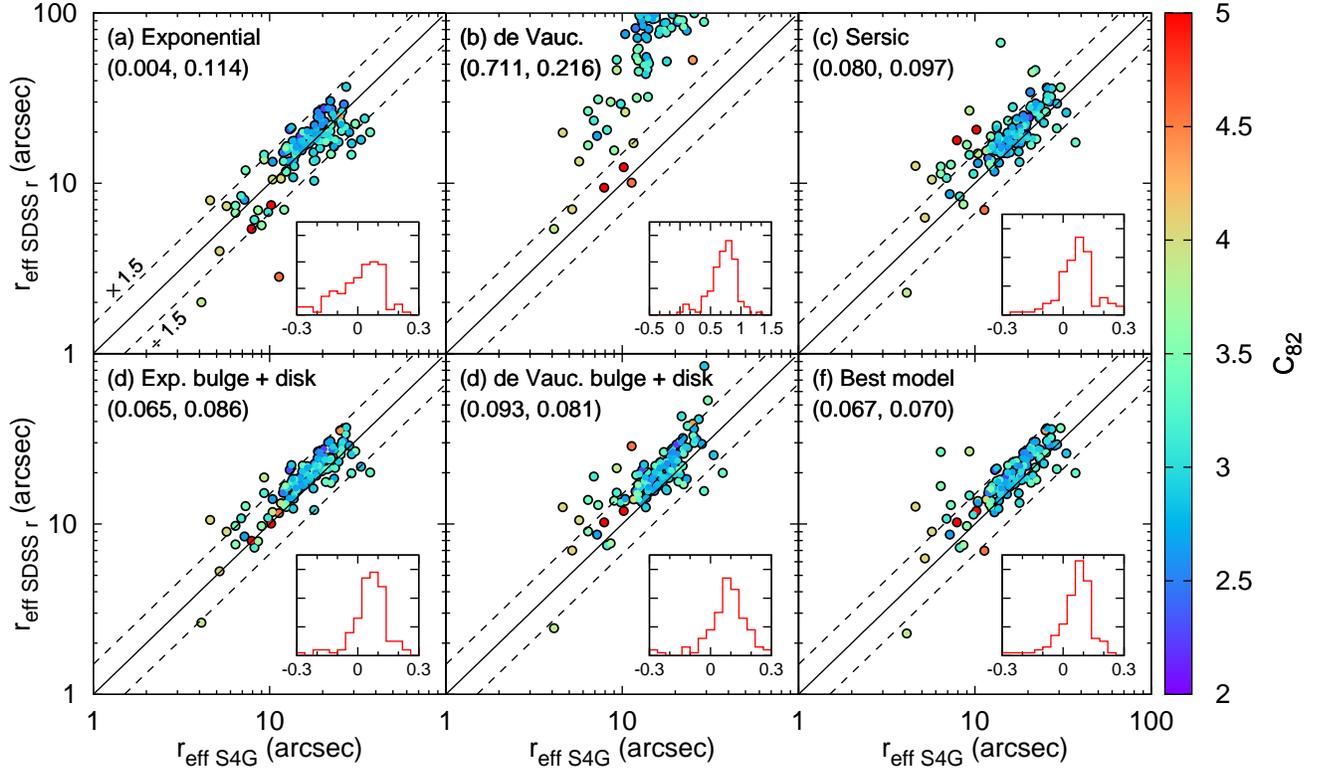}}\\
\caption{SDSS $r$-band effective radii from \cite{Lackner:2012}
  compared to the \sfg\ ones at 3.6\,$\micron$. The Lackner radii
  result from fitting each galaxy with various 2D models:
  (a) single exponential profile; (b) single de Vaucouleurs profile
  (the full range is not shown to avoid unnecessarily rescaling the
  axes of the other panels); (c) single S\'{e}rsic profile; (d)
  exponential bulge $+$ exponential disk; (e) de Vaucouleurs bulge $+$
  exponential disk; (f) model that best describes each galaxy. Points
  are color-coded according to the 3.6\,$\micron$ concentration index
  $C_{82}$. The small histograms show the distribution of
  $\log(r_\mathrm{eff\,SDSS\,r}/r_\mathrm{eff\,S4G})$. The mean and
  1$\sigma$ values of these distributions are quoted in dex units as
  $(\mu,\sigma)$ in each panel.\label{Lackner}}
\end{center}
\end{figure*}

We compare their effective radii along the major axis\footnote{For the
  bulge $+$ disk models, \cite{Lackner:2012} do not provide global
  $r_\mathrm{eff}$ for the whole galaxy, only for each individual
  component. We therefore added up the bulge and disk 1D profiles
  along their respective major axes and computed a global
  $r_\mathrm{eff}$.} with the \sfg\ ones in Fig.~\ref{Lackner}. Panels
(a) to (e) show the five models described above, and panel (f) shows
the model deemed by \cite{Lackner:2012} as the most representative for
each particular galaxy. The $r$-band radii are systematically larger
than the 3.6\,$\micron$ ones by $\sim 0.07-0.08$\,dex. This is
quantitatively consistent with the wavelength dependence of
$r_\mathrm{eff}$ found by \cite{Lange:2014}, due to color
gradients. After accounting for this offset, in term of scatter the
``best model'' option in panel (f) provides the best agreement with
our non-parametric sizes, with a 1$\sigma$ scatter of merely $\pm
0.07$\,dex ($\pm 16$\%). The bulge$+$disk models yield a similar
accuracy. The S\'{e}rsic fit does a somewhat poorer job, recovering
the non-parametric sizes within $\pm 22$\%. For the single exponential
fit, the scatter around the 1:1 line is broader and more skewed, as
such model is only suitable for almost bulgeless systems. Finally,
the de Vaucouleurs profile completely overestimates the sizes except
for some compact objects.

\section{Ancillary data and online access}\label{sec_irsa}
The enhanced data products from our different pipelines are available
through a dedicated webpage at the NASA/IPAC Infrared Science
Archive\footnote {\tt
  http://irsa.ipac.caltech.edu/data/SPITZER/S4G/}. For each galaxy,
users can download the P1 images and weight-maps, the P2 masks, the P3
radial profiles, the P4 GALFIT decompositions, and the P5 stellar mass maps.

IRSA also hosts an interactive catalog that contains all the global
properties measured in P3 for each galaxy: central coordinates,
background level and noise, outer isophotal sizes and shapes, apparent
and absolute asymptotic magnitudes, stellar masses, and concentration
indices. We have also incorporated into this catalog other parameters
mined from external sources. In particular, the catalog contains
redshift-independent distances from NED, and a variety of ancillary
measurements compiled in HyperLeda such as optical sizes, magnitudes
and colors, radial velocities, gas content and internal
kinematics.\footnote{Note that our IRSA catalog is not dynamically
  linked to NED or HyperLeda. Therefore, changes in these external
  sources will not be reflected in the values quoted in the IRSA
  catalog.} This allows users to define detailed subsamples within
\sfg\ according to different selection criteria; the resulting list of
galaxies can be then fed into the IRSA query system to retrieve the
images, profiles and other data for the selected objects.

\section{Summary}\label{sec_summary}
\sfg\ is the largest and most homogeneous inventory of the
stellar mass and structure in the nearby universe. Our 3.6\,$\micron$
and 4.5\,$\micron$ data peer through dust at the old stellar backbone
of more than 2300 nearby galaxies of all morphological types in
different environments.

In this paper we have described the methodology and algorithms of two
of the pipelines in our data flow. Pipeline 2 creates masks of
foreground stars and background galaxies. Pipeline 3 measures the
background level and noise, performs surface photometry on the images,
and derives global quantities such as asymptotic magnitudes, stellar
masses, isophotal sizes and shapes, and concentration indices.

We find a structural transition between disk-dominated to
bulge-dominated galaxies at stellar masses between $10^{10}$ and
$10^{11}\,M_{\sun}$, concurrent with a drop in the specific SFR from
$\sim10^{-10.5}\,\mathrm{yr}^{-1}$ to
$\sim10^{-11.5}\,\mathrm{yr}^{-1}$ with increasing
concentration. However, we find that not all galaxies with high
stellar concentration indices are quiescent spheroidal systems. Bars
and mergers can yield central concentrations of stellar mass while at
the same time enhancing the overall SFR of the galaxy. Extended UV
disks can also increase the SFR in the outer disk of galaxies that
look otherwise compact in the IRAC bands.

We have also studied the local stellar-mass size relation at
3.6\,$\micron$, which is consistent with previous results, but has the
advantage of being little influenced by gradients in dust extinction
or stellar age and metallicity. Early-type galaxies present an
approximately constant stellar mass surface density of
$10^9\,M_{\sun}\,\mathrm{kpc}^{-2}$. In late-type galaxies the surface
density decreases with mass from $10^9\,M_{\sun}\,\mathrm{kpc}^{-2}$
to $10^7\,M_{\sun}\,\mathrm{kpc}^{-2}$.

Given the large ancillary value of \sfg, we have made our enhanced
data products available to the community through a dedicated webpage
at IRSA. This data release includes not only the products described in
the present paper (masks, radial profiles and derived global
quantities), but also other enhanced products from our various
pipelines, including science-ready mosaics and weight-maps, 2D image
decompositions, and stellar mass maps.

\newpage
\acknowledgements

The authors are grateful to the entire \sfg\ team for their collective
effort in this project. We also thank the staff at IRSA, and in
particular Justin Howell, for implementing the online access to our
data. We acknowledge useful suggestions from an anonymous referee,
which helped to improve the scientific content of this paper. We thank
Rebecca Lange for sharing the contour data of the GAMA mass-size
relation.

JCMM acknowledges the receipt of an ESO Fellowship. This work was also
co-funded by NASA JPL/Spitzer grant RSA 1374189 provided for the \sfg\
project. JCMM, KS and TK also acknowledge support from the National
Radio Astronomy Observatory, which is a facility of the National
Science Foundation operated under cooperative agreement by Associated
Universities, Inc. We also acknowledge financial support from the
DAGAL network from the People Programme (Marie Curie Actions) of the
European Union’s Seventh Framework Programme FP7/2007-2013/ under REA
grant agreement number PITN-GA-2011-289313. EA and AB also acknowledge
financial support from the CNES (Centre National d'Etudes Spatiales -
France). J.H.K. acknowledges financial support from the Spanish MINECO
under grant number AYA2013-41243-P. A.G.d.P. acknowledge financial
support under grant number AYA2012-30717.

This work is based on data acquired with the Spitzer Space Telescope,
and makes extensive use of the NASA/IPAC Extragalactic Database (NED),
both of which are operated by the Jet Propulsion Laboratory,
California Institute of Technology under a contract with the National
Aeronautics and Space Administration (NASA). We have also made use of
the HyperLeda database (http://leda.univ-lyon1.fr) and NASA's
Astrophysics Data System Bibliographic Services.

{\it Facilities:} \facility{{\it Spitzer}}

\appendix
\section{Pipeline 1 technical details}\label{App_P1}
The \sfg\ Pipeline 1 is an evolved version of the IRAC pipeline used
for the SINGS Legacy Program (\citealt{Kennicutt:2003}). A major
change to the code from previous versions is the removal of a step to
improve the registration of individual frames using cross-correlation,
because the Spitzer calibration now uses 2MASS to improve the
astrometry. Also, sky subtraction has been moved into Pipeline
3. Below we summarize the steps in the pipeline.

\subsection{Saturation removal and rough cosmic ray rejection}
The first step is optional and only occurs when the IRAC observations
were taken in High Dynamic Range (HDR) mode. In this mode a short
exposure is taken before the main exposure at each location. For the
vast majority of the observations in the sample, the main observation
is 30 seconds and the HDR exposure is 1.2 seconds. For new galaxies
that were only observed during the warm mission we did not use HDR
mode since very few have bright nuclei.

If the Basic Calibrated Data (BCD) images have an HDR counterpart and
the flux in the pixel is above $10^6$\,MJy\,sr$^{-1}$, the value from
the normal BCD is replaced by the HDR one. In addition, we do a rough
cosmic ray removal by replacing the value of the pixels in the normal
BCD image with the value in the HDR pixel when the difference is above
3\,MJy\,sr$^{-1}$.

\subsection{Optical distortion correction and rotation}
Before we can compare the sky level in overlapping BCD images, we must
first correct each BCD for optical distortion and rotate the images to
have the same orientation. This is required to get pixels to all have
the same projeted size and orientation on the sky. To do this we use
the Drizzle routine and create a new optically corrected BCD image for
each BCD image in the mosaic. Once we have all the images with
constant size on the sky we can find the matching regions of overlap.

\subsection{Determining the background offset between BCD images}
For all pairs of overlapping BCD images where at least a minimum
number of pixels (nominally 20,000) overlap between the two images we
find the background offset. To determine the background offset we find the
difference in the 20th percentile pixel flux in both images. Using the
20th percentile has several advantages over a mean or median. Our goal
is to find the change in the background level rather than any changes
when pixels receive flux from resolved sources. Sub-pixel changes in
the telescope position will change the flux level of pixels observing
a point source substantially with the IRAC undersampled optics. These
same sub-pixel changes do not affect the observations of the
unresolved background. Also, some detector artifacts such as mux bleed
and column pull down can affect large areas leading to biases in even
the median, while using the 20th percentile value minimizes the bias.

For some archive observations the original frames had minimal overlap
for the mosaic.  In those cases we had to reduce the required number
of pixels in the overlap region to be as low as 5000.

\subsection{Solving for the background offset correction}
Once we have the background offset between each pair of overlapping
BCD images, we used the method of \cite{Regan:1995} to find the
background offset for each image that minimizes the residuals. This
method performs a least square minimization of the residuals using a
single offset that is applied to each image.

Due to their small size, many sample galaxies were not observed in
IRAC mapping mode but instead were observed in dither mode. In this
case two (or even three) non-overlapping regions of the sky are
observed and there is a need of additional information to obtain a
solution. Therefore, when the regions of the sky do not overlap we add
in an offset between a BCD image in each region of the sky. This will
force the sky level to be similar between the non-overlapping regions
of the sky.

The set of measured offsets of overlapping images does not provide a
zero point for the solution, which requires having an external
paramenter fixed to set the zero point for the relative offsets. To
minimize the effect of applying the offsets on the background level,
we place the additional constraint that the average offset should have
a value of 0.  Once we know the correction that needs to be applied,
we create a corrected version of each input BCD image.

\subsection{Drizzle the corrected images and update FITS header}
The final mosaics are created using the standard STDAS PYRAF task
Drizzle. The resuling images have a plate scale of 0.75\arcsec\ per
pixel and are oriented to have north to the top. We have to make a
correction to the output of the drizzling task to recover the correct
photometry. Drizzle assumes the pixels have units of flux instead of
surface brightness. We apply a correction to account for the smaller
pixel area in the final mosaic. Drizzle also returns a weight map that
shows the number of seconds of integration time for each final
pixel. This image is also included in the \sfg\ archive and is used by
subsequent \sfg\ pipelines.

In the final step we update the FITS header to include a list of the
Astronomical Observation Requests (AORs) that contributed to the final
mosaic and set all the pixels outside of the field of view to have the
value of NAN.

\end{document}